\documentclass[journal]{IEEEtran}
\ifCLASSINFOpdf
\else
\fi
\usepackage{graphicx}
\usepackage{setspace}
\usepackage{amsmath}
\usepackage{amssymb} 
\usepackage{caption}
\usepackage{subcaption}
\usepackage{bm} 
\usepackage{cite}
\usepackage{float}
\usepackage[usenames,dvipsnames]{pstricks}
\usepackage{epsfig}
\usepackage{pst-grad} 
\usepackage{pst-plot} 
\usepackage{tabularx}
\usepackage{flushend}

\usepackage[T1]{fontenc}
\newcommand{\lb} {\left}
\newcommand{\rb} {\right}
\newcommand{\nn} {\nonumber}
\usepackage{xcolor}
\usepackage{lipsum}

\allowdisplaybreaks

\allowdisplaybreaks
\flushend
\usepackage[utf8]{inputenc}

 \IEEEaftertitletext{\vspace{-2.5\baselineskip}}
\begin{document}
\bstctlcite{IEEEexample:BSTcontrol}
\title{Transmitter Selection for Secrecy in Frequency-Selective Fading with Multiple Eavesdroppers and Wireless Backhaul Links}
\author{
  \IEEEauthorblockN{Shashi Bhushan Kotwal\IEEEauthorrefmark{1}, Chinmoy Kundu\IEEEauthorrefmark{2}, 
  Sudhakar Modem\IEEEauthorrefmark{3},
  Ankit Dubey\IEEEauthorrefmark{4}, 
  and Mark F. Flanagan\IEEEauthorrefmark{5}
  }
  \IEEEauthorblockA{\IEEEauthorrefmark{1}\IEEEauthorrefmark{3}\IEEEauthorrefmark{4}Department of EE, Indian Institute of Technology Jammu, Jammu \& Kashmir, India }
  \IEEEauthorblockA{\IEEEauthorrefmark{2}\IEEEauthorrefmark{5}School of Electrical and Electronic Engineering, University College Dublin, Belfield, Ireland}
  \textrm{
  {{\IEEEauthorrefmark{1}sbkotwal@ieee.org},
  \{\IEEEauthorrefmark{3}sudhakar.modem ,\IEEEauthorrefmark{4}ankit.dubey\}
  @iitjammu.ac.in}, {\IEEEauthorrefmark{2}chinmoy.kundu@ucd.ie}, 
  {\IEEEauthorrefmark{5}mark.flanagan@ieee.org}}}
  \author{Shashi Bhushan Kotwal,~\IEEEmembership{Student Member,~IEEE,} Chinmoy Kundu,~\IEEEmembership{Member,~IEEE,} Sudhakar Modem,~\IEEEmembership{Member,~IEEE,} and Mark F. Flanagan,~\IEEEmembership{Senior Member, IEEE}

\thanks{Shashi Bhushan Kotwal and Sudhakar Modem are with Indian Institute of Technology Jammu, India
(email: \{shashi.kotwal, sudhakar.modem\}@iitjammu.ac.in).}
\thanks{Chinmoy Kundu and Mark F. Flanagan are with University College Dublin, Ireland (email: chinmoy.kundu@ucd.ie, mark.flanagan@ieee.org).}
\thanks{This publication has emanated from research supported in part by Science Foundation Ireland (SFI) under Grant Number 17/US/3445 and is co-funded under the European Regional Development Fund under Grant Number 13/RC/2077, and in part by the Science and Engineering Research Board, India sponsored project ECR/2018/002795.
} 

}
\maketitle
\thispagestyle{empty}
\pagestyle{plain} 
\begin{abstract}
This paper investigates the secrecy performance of sub-optimal transmitter selection (SS) and optimal selection (OS) schemes in a system comprising multiple transmitters, multiple eavesdroppers, and a single destination in frequency-selective fading channels with single carrier cyclic prefix modulation. Considering unreliable backhaul links between the access point and the transmitters, we analyze secrecy performance in two scenarios: when backhaul activity knowledge is available (KA) and when backhaul activity knowledge is unavailable (KU). Closed-form expressions for the secrecy outage probability (SOP) and ergodic secrecy rate (ESR) are derived. We also provide the corresponding asymptotic expressions when the backhaul links are unreliable as well as when the backhaul links are active in both KA and KU scenarios. We show that in the backhaul KA scenario, the asymptotic SOP and the slope of the ESR are governed by the number of transmitters and the backhaul reliability factor. In contrast, only the backhaul reliability factor influences these two parameters in the backhaul KU scenario. These parameters are independent of the number of eavesdroppers in both backhaul KA and KU scenarios. We also observe that the secrecy diversity order of both selection schemes is identical when all backhaul links are active. 

\end{abstract}
\section{Introduction}\label{sec_intro}

Future dense heterogeneous networks in beyond 5G, such as the Internet-of-Things (IoT), are envisaged to be comprising low-complexity small-cell networks \cite{context_aware_ieee_mag_22,  6G_applications_OJ_2020, ge_2014_5g, backhaul_Yin, KIM_PLS_Nakagami_TSP_2017, Kim_2016_CPSC_Globecom,  Kim_PLS_dACDD_TC_2020, SBK_VTC21, Kim_2016_CPSC_Trans, other_2017_CPSC, Kim_2018_CDD_JOUR}. In such networks, providing data security through physical layer security (PLS) techniques is considered a suitable choice due to its relatively simple channel coding techniques \cite{Kim_2016_CPSC_Globecom, Kim_PLS_dACDD_TC_2020, Chinmoy_TVT_2019, SBK_VTC21}. Improving PLS by selecting nodes (antenna, transmitter, relay, and or destination) which improves diversity gain has been studied extensively in \cite{Kim_2016_CPSC_Globecom, Kim_PLS_dACDD_TC_2020, Chinmoy_TVT_2019, SBK_VTC21, Chinmoy_TWC_2015, Mallik_TWC_14, Shilpa_ICCST19, Zhao_MIMO_outdated_CSI_TC18,  Chinmoy_GC16, Shalini_GC20, Chinmoy_letters21, Kim_2016_CPSC_Trans,other_2017_CPSC,Kim_2018_CDD_JOUR, ESR_SBK_OS_Multi_D_Unpub, Kim_2015_CPSC}. 
In selecting nodes, the availability of the global channel state information (CSI) can facilitate the optimal selection (OS) scheme which has been considered in \cite{Chinmoy_TVT_2019, SBK_VTC21, Chinmoy_TWC_2015}. The absence of either destination or eavesdropper CSI generally leads to sub-optimal selection (SS) schemes \cite{Chinmoy_TWC_2015, Kim_2016_CPSC_Globecom,   Kim_PLS_dACDD_TC_2020, ESR_SBK_OS_Multi_D_Unpub, Kim_2016_CPSC_Trans, other_2017_CPSC, Kim_2018_CDD_JOUR, SBK_VTC21,Chinmoy_TVT_2019}. 

Small-cell networks require backhaul links between the small-cell transmitters and the base station or control unit for connectivity \cite{ KIM_PLS_Nakagami_TSP_2017, Kim_2016_CPSC_Globecom}. Due to the low cost and deployment flexibility, wireless backhauls are preferred  \cite{ge_2014_5g, KIM_PLS_Nakagami_TSP_2017}. However, the inherently random nature of the wireless medium makes the backhauls unreliable \cite{KIM_PLS_Nakagami_TSP_2017, Kim_2016_CPSC_Globecom, backhaul_Yin}.
The available knowledge of backhaul activity makes it possible to select transmitters with active backhaul links only. As a result, the secrecy performance can be improved \cite{SBK_VTC21}. We refer to this scenario as the backhaul knowledge available (KA) scenario as in \cite{SBK_VTC21}.
However, it is not always possible to acquire knowledge of the backhaul activity \cite{Chinmoy_TVT_2019}, we  refer to this scenario as the backhaul knowledge unavailable (KU) scenario. In the backhaul KU scenario, the selection of transmitters is independent of the backhaul activity. 

In addition to the backhaul activity knowledge,
the secrecy performance is also influenced by the channel fading \cite{Chinmoy_TVT_2019, Shalini_GC20, Chinmoy_letters21,Kim_2016_CPSC_Globecom,   Kim_2016_CPSC_Trans, SBK_VTC21, Kim_PLS_dACDD_TC_2020}. Transmitter selection for secrecy considering wireless backhauls has been studied in the context of flat fading  in \cite{Chinmoy_TVT_2019, Shalini_GC20, Chinmoy_letters21}, and in the context of frequency-selective fading  in \cite{Kim_2016_CPSC_Globecom, SBK_VTC21, Kim_PLS_dACDD_TC_2020, Kim_2016_CPSC_Trans, other_2017_CPSC, Kim_2018_CDD_JOUR}. In a frequency-selective fading channel, the use of single carrier-cyclic prefix (SC-CP) signaling  improves the link capacity compared to orthogonal frequency division multiplexing (OFDM) \cite{M_Path_Div_Gain_conf_09}. The improvement in link capacity is attributed to the SC-CP signaling 
which is able to achieve multi-user and multi-path diversity. 

Leveraging SC-CP signaling in the frequency-selective fading channel, the authors improved the secrecy performance of small-cell networks with wireless backhaul in \cite{Kim_2016_CPSC_Globecom, Kim_2016_CPSC_Trans, other_2017_CPSC, Kim_2018_CDD_JOUR, SBK_VTC21, Kim_PLS_dACDD_TC_2020}. 
Among these,  the secrecy outage probability (SOP) was improved using the SS scheme in the backhaul KA scenario in \cite{Kim_2016_CPSC_Globecom,   Kim_PLS_dACDD_TC_2020}. Whereas, the authors improved the SOP using the OS scheme in both the backhaul KA and KU scenarios in \cite{SBK_VTC21}.   The literature \cite{Kim_2016_CPSC_Globecom,  Kim_PLS_dACDD_TC_2020, SBK_VTC21} studied secrecy in the presence of a single eavesdropper only. Moreover, the ESR analysis was not done in \cite{Kim_2016_CPSC_Globecom,  Kim_PLS_dACDD_TC_2020, SBK_VTC21}. 

The ESR performance was considered in the presence of multiple eavesdroppers using the SS scheme in \cite{Kim_2016_CPSC_Trans, other_2017_CPSC, Kim_2018_CDD_JOUR}. Only the backhaul KA scenario was considered. The ESR analysis of the SS scheme in the backhaul KU scenario and the OS scheme in both the backhaul KA and KU scenarios was not performed in \cite{Kim_2016_CPSC_Trans, other_2017_CPSC, Kim_2018_CDD_JOUR}. While the ESR analysis of the SS and OS scheme was performed in the frequency-selective fading channel in \cite{ESR_SBK_OS_Multi_D_Unpub}, the effect of backhauls and multiple eavesdroppers was not explored. 
In the presence of multiple eavesdroppers, 
 the ESR analysis was performed for the OS scheme in \cite{Chinmoy_letters21}. However, frequency-selective fading was not considered, and only the backhaul KA scenario was considered.

Applications developed for future networks require high-speed data transmission \cite{context_aware_ieee_mag_22,6G_applications_OJ_2020}.
High-speed data transmission often encounters frequency-selective fading in practice \cite{Book_WCOM_2016}. 
To the best of the authors' knowledge, under frequency-selective fading with SC-CP transmission and unreliable backhauls (KA and KU scenarios), the SOP and ESR for the OS scheme in the presence of multiple eavesdroppers are not available in the literature. For such a system, the SOP and ESR analyses for the SS scheme in the backhaul KU scenario have also not yet been studied. In \cite{ESR_SBK_OS_Multi_D_Unpub}, it was demonstrated that increasing the number of transmitters is beneficial for improving the ESR performance as compared to increasing the number of destinations by the same amount; however, only a single eavesdropper was considered. 

Motivated by the above discussion, in this paper we consider a system comprising multiple transmitters, a single destination, and multiple eavesdroppers under frequency-selective fading with SC-CP transmission. We then implement the SS and OS transmitter selection schemes to improve the secrecy of the network. The major contributions of the present work are as follows:


\begin{itemize}
      \item We derive the exact closed-form expression of the SOP and the ESR for the OS and the SS scheme in presence of multiple eavesdroppers considering the unreliable backhauls (in both the backhaul KA and KU scenarios) in frequency-selective fading channels with SC-CP signaling. 
     \item We present simplified asymptotic expressions of the SOP and ESR in the high-SNR regime to reduce the number of computations and find useful insights. We also provide secrecy diversity orders for both SS and OS schemes when all backhauls are active. We show that 
the number of eavesdroppers does not influence the asymptotic SOP limit and the slope of the asymptotic ESR.
     
     

     \item We introduce a generalized method to integrate the backhaul reliability factor into the secrecy performance analysis for the backhaul KA and KU scenarios 
     through the cumulative distribution function (CDF) of the ratio of the destination to eavesdropper link SNR in a unified manner. 
     
\end{itemize}



\textit{Notation:}  $||\mathbf{h}||$ denotes the Euclidean norm of a vector $\mathbf{h}$,  $|\mathcal{A}|$ denotes the size of set $\mathcal{A}$, $\mathbb{P[\cdot]}$ denotes the probability of an event, $\mathbb{E}\left[\cdot \right]$ denote the expectation operator, and $\left[x\right]^+=\max\{0,x\}$. $\binom{m}{n}$ denotes the binomial coefficient. $\Gamma(x)=\int_0^\infty t^{z-1}\exp(-t)dt$ denotes the complete Gamma function, while $\gamma(z,x)=\int_0^x t^{z-1}\exp(-t)dt$  and $\Gamma(z,x)=\int_x^\infty t^{z-1}\exp(-t)dt$ denote the lower and upper incomplete Gamma function, respectively. The probability density function (PDF) and the cumulative distribution function (CDF) of a random variable $X$ are denoted by $f_{X}(\cdot)$ and $F_{X}(\cdot)$, respectively. 
\section{System Model}
We consider a communication system comprising an access point A serving $K$ small-cell transmitters $\text{S}^{(k)}$ through backhaul links $b^{(k)}$ where $k\in\mathcal{S}=\{1,\ldots, K\}$. Access point A is analogous to a control unit in the dense heterogeneous small-cell networks \cite{Kim_2016_CPSC_Trans, SBK_VTC21}. The $K$ transmitters are transmitting data to a destination D in the presence of $N$ non-colluding passive eavesdroppers $\text{E}^{(n)}$, where $n\in\mathcal{N}=\{1,\ldots, N\}$.
Each node is equipped with a single antenna. We consider the frequency-selective fading model with SC-CP modulation. The destination channel $\mathbf{h}_{D}^{(k)}$ between $\text{S}^{(k)}$ and $\text{D}$ for each $k\in\mathcal{S}$ and the eavesdropping channel $\mathbf{h}_{E}^{(k,n)}$ between $\text{S}^{(k)}$ and $\text{E}^{(n)}$ for each $k \in \mathcal{S}$ and $n\in\mathcal{N}$ contain $M_{D}$ and $M_{E}$ paths, respectively. The links $\mathbf{h}_{D}^{(k)}\in\mathbb{C}^{1 \times M_{D}}$ for all $k \in \mathcal{S}$ are independent and identically distributed (i.i.d.) with element  ${h}_{D}^{(k)}(i)$ for each path $i \in\{1,\dots,M_{D}\}$  modeled as independent circularly symmetric complex Gaussian (CSCG) random variable (RV) with zero mean and unit variance. 
Similarly,  the links $\mathbf{h}_{E}^{(k,n)}\in\mathbb{C}^{1 \times M_{E}}$ for all $k\in\mathcal{S}$ and  $n\in\mathcal{N}$ are i.i.d. with ${h}_{E}^{(k,n)}(j)$ for each path  $j \in\{1,\dots,M_{E}\}$ modeled as an independent CSCG RV with zero mean and unit variance.
Each backhaul $b^{(k)}$ 
is modeled as an i.i.d. Bernoulli RV $\mathbb{I}^{(k)}$ with the probability of active and inactive backhaul indicated by  $\mathbb{P}\left[\mathbb{I}^{(k)}=1\right]=\zeta$ and  $\mathbb{P}\left[\mathbb{I}^{(k)} =0\right]=1-\zeta$, respectively \cite{Chinmoy_letters21}, where $0 \le \zeta \le 1$ is the backhaul reliability factor. 

As SC-CP modulation is considered, the SNR  $\gamma_{D}^{(k)}$ for the $\text{S}^{(k)}$-$\text{D}$ link and the SNR $\gamma_{E}^{(k,n)}$ for the $\text{S}^{(k)}$-$\text{E}^{(n)}$ link with active $b^{(k)}$ are written as \cite{Kim_2015_CPSC}
\begin{align}\label{gamma_Dk}
    \gamma_{D}^{(k)} = \frac{ a_{D}^{(k)} P_T }{{\sigma_{D}}^{2}} ||\mathbf{h}^{(k)}_{D}||^2
    ~\text{and}~
    \gamma_{E}^{(k,n)} = \frac{ a_{E}^{(k,n)} P_T }{{\sigma_{E}}^{2}} ||\mathbf{h}^{(k,n)}_{E}||^2, 
\end{align}
respectively, where $P_T$ is the power transmitted by $\text{S}^{(k)}$, and $a_D^{(k)}$ and $a_{E}^{(k,n)}$ are the path-loss factors for the $\text{S}^{(k)}$-$\text{D}$ and $\text{S}^{(k)}$-$\text{E}^{(n)}$ links, respectively. The average noise powers at $D$ and $E^{(n)}$ are ${\sigma_{D}}^{2}$ and  ${\sigma_{E}}^{2}$, respectively. The SNR $\gamma_{D}^{(k)}$ is shown to follow Gamma distribution with CDF \cite{Kim_2015_CPSC} 
\begin{align}\label{eq_CDF_D}
 F_{{\gamma}_{D}^{(k)}}(x) 
 &=1-\exp\big(-\frac{x}{\lambda_D}\big)\sum_{m_D=0}^{M_D-1}\frac{1}{m_D!}\big(\frac{x}{\lambda_D}\big)^{m_D},
\end{align}
  where the average SNR of the link is $\mathbb{E}[\gamma_D^{(k)}]=M_D\lambda_D$ and $\lambda_D$ is the average SNR per multipath component. From \eqref{eq_CDF_D}, it is apparent that the underlying frequency-selective channel upon SC-CP modulation behaves as a narrowband Nakagami-$m$ fading channel with integer shape parameter $m=M_{D}$ and scale parameter $M_{D}\lambda_{D}$ as also observed in \cite{ESR_SBK_OS_Multi_D_Unpub}.
  Thus, the analysis of this paper can also provide the results of the system considered here with flat Nakagami and Rayleigh fading channels. 
The CDF $F_{\gamma_{E}^{(k,n)}}(x)$ can be obtained by replacing ${\gamma}_{D}^{(k)}$ by ${\gamma}_{E}^{(k,n)}$, $\lambda_D$ by $\lambda_E$, and $M_D$ by $M_E$ in \eqref{eq_CDF_D}. Due to the non-colluding eavesdroppers, the secrecy performance corresponding to $\text{S}^{(k)}$ is measured against the eavesdropper with maximum SNR which is given by $\gamma_{E}^{(k)}=\max\limits_{n\in \mathcal{N}} \{\gamma_{E}^{(k,n)}\}$.
The CDF of $\gamma_{E}^{(k)}$ is given by 
\begin{align}
\label{eq_cdf_mult_eaves}
 F_{\gamma_{E}^{(k)}}(y)
&=\Big[1-\exp\big(-\frac{x}{\lambda_E}\big)\sum_{m_E=0}^{M_E-1}\frac{1}{m_E!}\big(\frac{x}{\lambda_E}\big)^{m_E}\Big]^N.
\end{align}

As we are evaluating the performance of transmitter selection schemes with unreliable wireless backhaul, by accounting for the backhaul reliability factor in the analysis, the secrecy rate corresponding to the selected transmitter $k^*$ for the $\text{S}^{(k^*)}$-$\text{D}$ link in bits per channel use (bpcu) can be defined as $C_{S}^{(k^*)} = \big[\log_2\big({\Gamma}_{S}^{(k^*)}\big)\big]^+$ \cite{Wyner_1975} with
\begin{align}\label{eq_Gamma_S_k_n_gen}   \hat{\Gamma}_{S}^{(k^*)}=\frac{1+\hat{\gamma}_{D}^{(k^*)}}{1+\hat{\gamma}_{E}^{(k^*)}},
\end{align}
where $\hat{\gamma}_{D}^{(k^*)}$ and $\hat{\gamma}_{E}^{(k^*)}$ are the SNRs at D and E, respectively, for the selected transmitter $k^*$ including backhaul reliability factor. The definition of $\hat{\gamma}_{D}^{(k^*)}$ and $\hat{\gamma}_{E}^{(k^*)}$ depend on the particular selection scheme and the  backhaul activity knowledge.  The CDF of $\hat{\Gamma}_{S}^{(k^*)}$ can be determined as
\begin{align}\label{eq_sop_to_CDF_conversion}
    F_{\hat{\Gamma}_{S}^{(k^*)}}(x)
    &=\int_0^\infty F_{\hat{ \gamma}_D^{(k^*)}}\left({x\left(1+y\right)-1}\right)f_{\hat\gamma_E^{(k^*)}}(y)dy.
\end{align}
Utilizing $F_{\hat{\Gamma}_{S}^{(k^*)}}(x)$, the SOP and the ESR for the transmitter selection schemes are evaluated in this paper.   
The SOP of the system corresponding to any transmitter selection scheme is defined as the probability that the secrecy rate $C_{S}^{(k^*)}$ is less than the threshold $R_{\text{th}}$ as
\begin{align}\label{eq_SOP_deinition}
    P_{out}=\mathbb{P}\lb\{\log_2\big(\hat{\Gamma}_{S}^{(k^*)}\big) \le R_{\text{th}}\rb\}
    =F_{\hat{\Gamma}_{S}^{(k^*)}}(2^{R_{\text{th}}}).
\end{align}
Further, the ESR of the system corresponding to any transmitter selection scheme is evaluated as
\begin{align}\label{eq_ESR_CDF_eqn}
 C_{\mathrm{erg}}  &= \mathbb{E}\big[C_{S}^{(k^*)}\big] = \frac{1}{\text{ln}(2)}\int_1^\infty \frac{1-F_{\hat{\Gamma}_{S}^{(k^*)}}(x)}{x} dx. 
\end{align}

We have generalized the secrecy performance analysis (SOP and ESR) methodology by utilizing \eqref{eq_sop_to_CDF_conversion} wherein the backhaul reliability factor is included in  $F_{\hat{\Gamma}_{S}^{(k^*)}}$ irrespective of the backhaul activity knowledge (KA or KU scenarios) and the selection scheme (SS or OS)\footnote{$F_{\hat{\Gamma}_{S}^{(k^*)}}(x)$ can be utilized to evaluate other performance metrics such as the probability of non-zero secrecy rate (PNZ), secrecy throughput \cite{Effective_Secrecy_throughput_ICC_2014}, etc. in the multiple eavesdroppers' system and for various node selection schemes including SS scheme (in \cite{Kim_2015_CPSC, Kim_2016_CPSC_Trans, Kim_2018_CDD_JOUR,other_2017_CPSC}), and the OS scheme.}.  The system model and the approach utilized in this paper are different from those of \cite{Kim_2016_CPSC_Globecom, Kim_2016_CPSC_Trans, other_2017_CPSC, Kim_2018_CDD_JOUR, SBK_VTC21, Kim_PLS_dACDD_TC_2020}. Further, due to the multiple eavesdroppers and unreliable backhaul links considered in this paper, which are not present in \cite{ESR_SBK_OS_Multi_D_Unpub}, the CDF of the ratio of SNRs in this paper is derived differently from that of \cite{ESR_SBK_OS_Multi_D_Unpub}; as a result, the SOP and ESR for the transmitter selection schemes in this paper cannot be derived from simple modifications of the results of \cite{ESR_SBK_OS_Multi_D_Unpub}. 

 
 In the following sections, we detail the analysis method of how KA and KU scenarios are incorporated in the secrecy performance analysis for both transmitter selection schemes (SS and OS).

\section{SOP: Backhaul KA Scenario}\label{section_sop_ka_scenario}
This section presents the closed-form SOP and its asymptotic limit for both the SS and OS transmitter selection schemes in the backhaul KA scenario. The 
KA scenario is one in which the knowledge of transmitters with active backhaul links is available. Therefore, the selection can be made among the transmitters with active backhauls. 

Following \eqref{eq_sop_to_CDF_conversion}, we need $F_{\hat{ \gamma}_D^{(k^*)}}$ and $f_{\hat\gamma_E^{(k^*)}}$ to evaluate the SOP in the KA scenario. For these distributions, we require distributions corresponding to the individual transmitter with the backhaul reliability factor taken into consideration.  With the backhaul reliability factor taken into consideration, the SNR at D  corresponding to $\text{S}^{(k)}$   is given by $\hat{\gamma}_D^{(k)} = \gamma_D^{(k)} \mathbb{I}^{(k)}$.
We note $\hat{\gamma}_D^{(k)}$ follows a mixture distribution of $\gamma_D^{(k)}$ and $\mathbb{I}^{(k)}$ with the CDF expressed as\cite{Kim_2016_CPSC_Globecom}
\begin{align}\label{eq_CDF_Dk_BHKA}
 F_{\hat\gamma_D^{(k)}}(x) =1-\zeta \exp\big(-\frac{x}{\lambda_D}\big)\sum_{m_D=0}^{M_D-1}\frac{1}{m_D!}\big(\frac{x}{\lambda_D}\big)^{m_D}.
\end{align}
An eavesdropper can intercept messages meant for the destination only if the backhaul of the selected transmitter is active; therefore, we take  $\hat\gamma_E^{(k)}=\gamma_E^{(k)}$ throughout the analysis. 

\subsection{SOP of the SS scheme}\label{subsection_sop_ss_ka}
The SS scheme is implemented in the absence of the CSI of the $\text{S}^{(k)}$-$\text{E}^{(n)}$ link. The SS scheme  selects a transmitter for which the main channel SNR at the destination is maximum. For the SS scheme,  $\hat{\Gamma}_{S}^{(k^*)}$ in
\eqref{eq_Gamma_S_k_n_gen} is defined as 
\begin{align}\label{eq_Gamma_sk_SS_BHKA}
\hat{\Gamma}_{S}^{(k^*)}=\frac{1+\max\limits_{k\in \mathcal{S}}\{\hat{\gamma}_{D}^{(k)}\}}{1+\hat{\gamma}_{E}^{(k^*)}}.
\end{align}
As link $\text{S}^{(k)}$-$\text{E}^{(k)}$ for each $k$ is i.i.d., the SNR $\hat{\gamma}_{E}^{(k^*)}$ follows the distribution of $\gamma_E^{(k)}$ as in \eqref{eq_cdf_mult_eaves}. 
The CDF  $F_{\hat{\Gamma}_{S}^{(k^*)}}(x)$ is then  evaluated as
\begin{align}\label{eq_CDF_GAMMA_S_SS_BHKA_int_form}
  &F_{\hat{\Gamma}_{S}^{(k^*)}}(x)
 =\int_0^\infty \lb[F_{\hat{ \gamma}_{D}^{(k)}}\left({x\left(1+y\right)-1}\right)\rb]^K f_{\hat{\gamma}_{E}^{(k^*)}}(y)dy\\
 &=1-\sum_{k=1}^{K}\sum_{n=0}^{N-1}(-1)^{k+1}\frac{\binom{K}{k}\binom{N-1}{n}\zeta^k N \exp\big(\frac{-k(x-1)}{\lambda_D}\big)  }{(\lambda_E)^{M_E}\Gamma(M_E)}\nn\\
   &\times \int_{0}^{\infty} \underbrace{\bigg(\sum_{m_D=0}^{M_D-1}\sum_{\mu=0}^{m_D}\sum_{l=0}^{m_D-\mu}\frac{(-1)^l\binom{m_D}{\mu}\binom{m_D-\mu}{l}x^{m_D-1}y^{\mu}}{(m_D!){\lambda_D}^{m_D}}\bigg)^k}_{\text{To be expanded using \eqref{eq_cross_mul_SS_BHKA}}}\nn\\
   &\times \underbrace{\bigg(\sum_{m_E=0}^{M_E-1}\frac{1}{m_E!}\left(\frac{y}{\lambda_E}\right)^{m_E}\bigg)^n}_{\text{To be expanded using \eqref{eq_cross_mul_SS_BHKA_2nd_time}}}y^{M_E-1}
   \nn\\
   &\times (-1)^n
   \exp\left(-\big(\frac{k \rho}{\lambda_D}+\frac{(n+1)}{\lambda_E}\big)y\right) dy\nn\\
 \label{eq_CDF_GAMMA_S_SS_BHKA_sol}
&=1-\sum_{k=1}^K\sum_{n=0}^{N-1} \sum_{\lb(\mathbf{m_D},\mathbf{\mu},\mathbf{l},\mathbf{m_E}\rb)\in\mathcal{X}} (-1)^{k+1} \binom{K}{k}\nn\\
&\times \zeta^k \Upsilon\frac{x^{\widehat{m}_{D}^{(k)} -\widehat{l}^{(k)}}\exp\big(-\frac{k x}{\lambda_D}\big)}{\big(x+\frac{(n+1)\lambda_D}{k \lambda_E}\big)^{\theta^{(k,n)}}},
\end{align}  
where
\begin{align}\label{eq_Upsilon_SS_BHKA}
    &\Upsilon \triangleq  (-1)^{n}\binom{N-1}{n}
   \bigg(\prod\limits_{q=1}^{k}(-1)^{l^{(q)}}\binom{m_{D}^{(q)}}{\mu^{(q)}}\binom{m_{D}^{(q)}-\mu^{(q)}}{l^{(q)}}\nn\\
   &\times \frac{ {(\lambda_D)}^{\mu^{(q)}-m_{D}^{(q)}}}{{(k)}^{\mu^{(q)}} m_{D}^{(q)}!}\bigg)\frac{N \exp\big(\frac{k}{\lambda_D}\big){(\lambda_D)}^{M_E+\widehat{m}_{E}^{(n)}}\Gamma\big(\theta^{(k,n)}\big)}{{(k\lambda_E)}^{M_E+\widehat{m}_{E}^{(n)}}\Big(\prod\limits_{i=1}^{n}m_{E}^{(i)}\Big){\Gamma(M_E)}},
\end{align}
and $\widehat{m}_{D}^{(k)}\triangleq  \sum_{q=1}^{k}{{m}_{D}}^{(q)}$, $\widehat{l}^{(k)}\triangleq  \sum_{q=1}^{k}{l}^{(q)}$,  $\widehat{\mu}^{(k)} \triangleq  \sum_{q=1}^{k}{\mu}^{(q)}$, $\widehat{m}_{E}^{(n)}\triangleq  \sum_{q=1}^{n}{{m}_{E}}^{(q)}$, and $
   \theta^{(k,n)}\triangleq  M_E+ \widehat{\mu}^{(k)}+\widehat{m}_{E}^{(n)}$.  We obtain the expression in \eqref{eq_CDF_GAMMA_S_SS_BHKA_sol} using the conversion of product-of-sums into a sum-of-products as
\begin{align}\label{eq_cross_mul_SS_BHKA}
&\Big(\sum_{i=0}^{\tau}\sum_{j=0}^i \sum_{u=0}^{i-j} f(i,j,u)\Big)^{\kappa}
=\nn\\
&\sum_{\mathbf{m_D} \in \mathcal{M}_{D}^{(\kappa)}} \sum_{\mathbf{\mu} \in \mathcal{U}^{(\kappa)}(\mathbf{m_D})}\sum_{\mathbf{l}\in\mathcal{L^{(\kappa)}(\mathbf{m_D,\mu})}}
\prod\limits_{q=1}^{{\kappa}}f(m_{D}^{(q)},\mu^{(q)},l^{(q)}),\\
\label{eq_cross_mul_SS_BHKA_2nd_time}
&\Big(\sum_{i=0}^{\mathcal{M}}f(i)\Big)^{\eta}
     =\sum_{\mathbf{m_E} \in \mathcal{M}^{(\eta)}} \prod_{p=1}^{{\eta}}f(m_{E}^{(p)}), 
\end{align}
and the integral solution of the form \cite[eq. (3.351.3)]{ryzhik_2007}.
In \eqref{eq_cross_mul_SS_BHKA},   $\mathcal{M}_{D}^{(\kappa)}$ is the set of integer vectors $[m_{D}^{(1)},\ldots,m_{D}^{(\kappa)}]$ containing $\kappa$ elements such that $m_{D}^{(q)}\in\{0 ,\ldots, \tau\}$ for each $q\in\{1 ,\ldots, \kappa\}$, $\mathcal{U}^{(\kappa)}(\mathbf{m_D})$ is the set of index vectors $[\mu^{(1)},\ldots,\mu^{(\kappa)}]$ containing $\kappa$ elements such that $\mu^{(q)} \in \{0,\ldots,m_{D}^{(q)}\}$, and  $\mathcal{L}^{(\kappa)}(\mathbf{m_D,\mu})$ is the set of integer vectors $[l^{(1)},\ldots,l^{(\kappa)}]$ containing $\kappa$ elements such that $l^{(q)} \in \{0,\ldots,(m_{D}^{(q)}-\mu^{(q)})\}$. In \eqref{eq_cross_mul_SS_BHKA_2nd_time}, $\mathcal{M}^{(\eta)}$ is the set of integer vectors $[m_{E}^{(1)},\ldots,m_{E}^{(\eta)}]$ containing $\eta$ elements such that $m_{E}^{(q)}\in\{0,\ldots, (M_E-1)\}$ for each $q\in\{1 ,\ldots, \eta\}$. In \eqref{eq_CDF_GAMMA_S_SS_BHKA_sol}, we denote the set of vector tuples $(\mathbf{m_D},\mathbf{\mu},\mathbf{l},\mathbf{m_E})$ by $\mathcal{X}$.   
The closed-form expression for the SOP can easily be obtained by substituting \eqref{eq_CDF_GAMMA_S_SS_BHKA_sol} into \eqref{eq_SOP_deinition}.


From \eqref{eq_CDF_GAMMA_S_SS_BHKA_sol} we observe that the SOP depends on $\zeta$ along with the other system parameters $K$, $N$, $M_D$, $M_E$, $\lambda_D$, and $\lambda_E$. The presence of multiple interdependent summations in \eqref{eq_CDF_GAMMA_S_SS_BHKA_sol} makes it difficult to understand how these parameters affect the SOP. Increasing $K$, $N$, and $M_D$ significantly increases the number of summation and multiplication terms, making the SOP evaluation computationally intensive. 
To identify how the parameters affect the system design, the expression of the SOP needs to be simple. In online systems, such as delay-sensitive IoT applications, where latency is a limiting factor, simplified system design is also preferred. Thus, the asymptotic analysis in the following section aims to simplify the SOP expression and find useful insights.

\subsection{Asymptotic analysis of the SS scheme}\label{subsection_asymp_sop_ss_ka}
For the asymptotic analysis of the SOP, we assume $\lambda_D \rightarrow{\infty}$ for a given $\lambda_E$ in this section and in the subsequent sections where  asymptotic SOP is derived \footnote{The asymptotic analysis in \cite{SBK_VTC21} is different to that of this paper. In \cite{SBK_VTC21}, the analysis was carried out by assuming high transmit power in a single eavesdropper system. However, we assume high received average SNR per multipath component at the destination for a given SNR at the eavesdroppers in a multiple eavesdropper scenario.}. 
This assumption is valid when the quality of the $\text{S}^{(k)}$-$\text{D}$ link is much better than any of the eavesdroppers. Using the Taylor series expansion of $\exp(-\frac{x}{\lambda_D})$  in \eqref{eq_CDF_Dk_BHKA}, we obtain
\begin{align}\label{eq_CDF_D_BHKA_Asymp_start}
  F_{\hat{\gamma}_{D}^{(k)}}(x) 
&=1-\zeta+ \sum_{i=1}^{\infty}\sum_{m_D=0}^{M_D-1}\frac{(-1)^i}{(m_D!)(i!)}\Big(\frac{x}{\lambda_D}\Big)^{i+m_D}.
\end{align}
In \eqref{eq_CDF_D_BHKA_Asymp_start}, we obtain $\lim\limits_{\lambda_D \rightarrow{\infty}}F_{\hat{\gamma}_{D}^{(k)}}(x)=1-\zeta$ and  by substituting it in \eqref{eq_CDF_GAMMA_S_SS_BHKA_int_form}, the asymptotic SOP is obtained as
\begin{align}\label{eq_SOP_Asympt_SS_BHKA}
  P_{out}^{\infty}=\left(1-\zeta\right)^K.
\end{align}
From \eqref{eq_SOP_Asympt_SS_BHKA}, we note that asymptotic SOP requires very few computations compared to the exact SOP expression in \eqref{eq_CDF_GAMMA_S_SS_BHKA_sol}.
It can be seen that the asymptotic SOP does not improve with SNR and saturates to a constant value governed by $\zeta$ and $K$ only. 
The asymptotic SOP in \eqref{eq_SOP_Asympt_SS_BHKA} approaches zero, when all backhaul links are active i.e., $\zeta=1$.  In this situation, \eqref{eq_SOP_Asympt_SS_BHKA} does not provide the rate of improvement of the SOP or its slope in the high-SNR regime. 
Therefore, we determine the slope of the SOP in the high-SNR regime in the following section through asymptotic analysis when $\zeta=1$. 

This observation in (\ref{eq_SOP_Asympt_SS_BHKA}) agrees with the SOP obtained for the SS scheme in  \cite{Kim_2018_CDD_JOUR}.  However, we have generalized the analysis approach to find both the SOP and the ESR of transmitter selection schemes (OS and SS) in the unreliable backhaul scenario by exploiting the distribution of $\hat{\Gamma}_{S}^{(k)}$. Moreover, the SOP of the system with multiple eavesdroppers in the backhaul KU scenario has not been studied in the literature. Further, we will also present the SOP analysis of the OS scheme in both the backhaul KA and KU scenarios in the paper, which is also not currently available in the literature.

\subsection{Asymptotic analysis of the SS scheme when $\zeta=1$}\label{section_div_order_ss}
We perform asymptotic analysis in this section following Section \ref{subsection_asymp_sop_ss_ka} when all backhaul links are active, i.e., $\zeta=1$.
We directly use $\zeta=1$ in \eqref{eq_CDF_D_BHKA_Asymp_start}. As $\lambda_D \rightarrow{\infty}$, we retain the lowest order term of $\frac{x}{\lambda_D}$ to obtain $\lim\limits_{\lambda_D \rightarrow{\infty}}{F_{{\gamma}_{D}^{(k)}}(x)} = \frac{1}{M_D!}\Big(\frac{x}{\lambda_D}\Big)^{M_D}$.
Applying the result in \eqref{eq_CDF_GAMMA_S_SS_BHKA_int_form}, the asymptotic SOP is evaluated as
\begin{align}\label{eq_SOP_Asympt_SS_BHKA_Perf_Bakchaul}
    &P_{out}^{\infty}
 =\sum_{\mu=0}^{K M_D}\sum_{n=0}^{N-1}\sum_{\mathbf{m}_{E} \in \mathcal{M}_{E}^{(n)}}(-1)^{n}\binom{K M_D}{\mu}\binom{N-1}{n}\nn\\
 &\times \frac{N(\lambda_E)^{\mu}{\rho}^\mu(\rho-1)^{K M_D-\mu}\Gamma\big(\varphi^{(n)}\big)}{{(\lambda_D)}^{K M_D}{(M_D!)}^K\Big(\prod\limits_{i=1}^{n}m_{E}^{(i)}!\Big) \Gamma(M_E){(n+1)}^{\varphi^{(n)}}},
\end{align}
where $\varphi^{(n)}\triangleq M_E+\mu+\widehat{m}_{E}^{(n)}$,  $\rho=2^{R_{\text{th}}}$, and  $\mathcal{M}_{E}^{(\mathbf{n})}$ is a set of index vectors $[m_{E}^{(1)} ,\ldots,  m_E^{(n)}]$ containing $n$ elements such that $m_{E}^{(i)} \in \{0 ,\ldots, (M_E-1)\}$ for each $i\in\{1 ,\ldots, n\}$.

To find the secrecy diversity order $d$ of the SS scheme from \eqref{eq_SOP_Asympt_SS_BHKA_Perf_Bakchaul}, we use the following definition \cite{DIV_GAIN_J_SEL_AREA_2013} 
\begin{align}\label{eq_div_order}
    d=  -\lim_{\lambda_D\rightarrow\infty}\frac{\ln {(P_{out})}}{\ln{(\lambda_D)}} =KM_D.
\end{align}
We note from (\ref{eq_div_order}) that the rate of improvement in the SOP depends equally on the number of transmitters and the multipath components in the main channel. This is due to the selection diversity and the diversity arising due to the SC-CP modulation.

\subsection{SOP of the OS scheme}\label{subsection_sop_os_ka}
In the OS scheme, a transmitter is selected for which the secrecy rate of the system is maximized. The OS scheme assumes that the CSI of the $\text{S}^{(k)}$-$\text{D}$ link as well as $\text{S}^{(k)}$-$\text{E}^{(n)}$ link for each $k\in\mathcal{S}$ and $n\in\mathcal{N}$ is available. Similar to Section \ref{subsection_sop_ss_ka}, we evaluate the SOP using \eqref{eq_SOP_deinition}. In the OS scheme, $\hat{\Gamma}_{S}^{(k^*)}$ in  \eqref{eq_Gamma_S_k_n_gen} is defined as
\begin{align}\label{eq_Gamma_sk_os_ka}
\hat{\Gamma}_{S}^{(k^*)}=\max_{k\in \mathcal{S}}\bigg\{\frac{1+\hat{\gamma}_{D}^{(k)}}{1+\hat{\gamma}_{E}^{(k)}}\bigg\}
,
\end{align}
where $\hat{\gamma}_{E}^{(k)}$ follows the distribution given by \eqref{eq_cdf_mult_eaves} as in (\ref{eq_Gamma_sk_SS_BHKA}). 
We need to find the distribution of $\hat{\Gamma}_{S}^{(k^*)}(x)$ for the performance analysis. Towards that goal, the CDF $F_{\hat{\Gamma}_{S}^{(k^*)}}(x)$ is evaluated as
\begin{align}\label{eq_CDF_GAMMA_S_OS_BHKA}
  &F_{\hat{\Gamma}_{S}^{(k^*)}}(x)  
      = \lb[\int_0^\infty F_{\hat{ \gamma}_{D}^{(k)}}\left({x\left(1+y\right)-1}\right)f_{\hat{\gamma}_{E}^{(k)}}(y)dy\rb]^K\\
  \label{eq_CDF_GAMMA_S_OS_BHKA_sol}
   &= \Bigg[1-\sum_{m_D =0}^{M_D-1}\sum_{\mu=0}^{m_D}\sum_{l=0}^{m_D-\mu}\sum_{n=0}^{N-1}\sum_{\mathbf{m_E}\in\mathcal{M}_{E}^{(n)}}(-1)^{n+l}\binom{m_D}{\mu}\nn\\
   &\times \binom{m_D-\mu}{l}\binom{N-1}{n}\frac{\zeta N {(\lambda_D)}^{\varphi^{(n)}-m_D}\Gamma(\varphi^{(n)})}{{(\lambda_E)}^{M_E+\widehat{m}_{E}^{(n)}}\Gamma(M_E)(m_D!)}\nn\\
   &\times \frac{x^{m_D-l}\exp\big(-\frac{x-1}{\lambda_D}\big)}{\Big(\prod\limits_{i=1}^{n}m_{E}^{(i)}!\Big){\big(x+\frac{(n+1)\lambda_D}{\lambda_E}\big)}^{\varphi^{(n)}}}\Bigg]^K.
\end{align}
We obtain the expression in \eqref{eq_CDF_GAMMA_S_OS_BHKA_sol} with the help of the general result in \eqref{eq_cross_mul_SS_BHKA_2nd_time} and using the integral solution of the form \cite[eq. (3.351.3)]{ryzhik_2007}. 
The closed-form expression of the SOP can be obtained easily from \eqref{eq_CDF_GAMMA_S_OS_BHKA_sol} using \eqref{eq_SOP_deinition}. Similar to \eqref{eq_CDF_GAMMA_S_SS_BHKA_sol}, the expression in \eqref{eq_CDF_GAMMA_S_OS_BHKA_sol} is computationally intensive. 


To obtain a simplified expression of the SOP and to identify the effect of system parameters on the SOP performance, we perform the asymptotic analysis of the SOP in the following section. 

\subsection{Asymptotic analysis of the OS scheme}\label{subsection_asymp_sop_os_ka}
With the assumption as in Section \ref{subsection_asymp_sop_ss_ka} for the asymptotic analysis, from \eqref{eq_CDF_Dk_BHKA} we obtain
$\lim\limits_{\lambda_D \rightarrow{\infty}}F_{\hat{\gamma}_{D}^{(k)}}(x)=1-\zeta$.  Then from \eqref{eq_CDF_GAMMA_S_OS_BHKA}, the asymptotic SOP is obtained as in \eqref{eq_SOP_Asympt_SS_BHKA} which is equal to that of the SS scheme. Hence, we conclude that the asymptotic SOP in the KA scenario does not depend on the particular selection scheme (SS or OS); only $\zeta$ and $K$ govern the asymptotic SOP.
Next, we proceed with the asymptotic analysis and secrecy diversity order analysis of the OS scheme when all backhauls are active.

\subsection{Asymptotic analysis of the OS scheme when $\zeta=1$}\label{section_div_order_os}
For the asymptotic analysis when all backhaul links are active, we obtain $\lim\limits_{\lambda_D \rightarrow{\infty}}{F_{{\gamma}_{D}^{(k)}}(x)} = \frac{1}{M_D!}\Big(\frac{x}{\lambda_D}\Big)^{M_D}$.
Using this in \eqref{eq_CDF_GAMMA_S_OS_BHKA} when $\zeta=1$, we evaluate the asymptotic SOP as
\begin{align}\label{eq_SOP_Asympt_OS_BHKA_Perf_Bakchaul}
  P_{out}^{\infty}
&=\bigg[\sum_{\mu=0}^{M_D}\sum_{n=0}^{N-1}\sum_{\mathbf{m_E}\in\mathcal{M}_{E}^{(n)}}(-1)^n \binom{M_D}{\mu}\binom{N-1}{n}\nn\\
&\times \frac{N {(\lambda_E)}^\mu \rho^\mu (\rho-1)^{M_D-\mu}\Gamma(\varphi^{(n)})}{{(\lambda_D)}^{M_D}(M_D!)\Big(\prod\limits_{i=1}^{n}m_{E}^{(i)}!\Big){(n+1)}^{\varphi^{(n)}}}\bigg]^K,
\end{align}
where $\mathcal{M}_{E}^{(\mathbf{n})}$ is defined as in \eqref{eq_SOP_Asympt_SS_BHKA_Perf_Bakchaul}. From \eqref{eq_SOP_Asympt_OS_BHKA_Perf_Bakchaul}, the secrecy diversity order $d$ is found to be the same as in  \eqref{eq_div_order}. This concludes that the diversity order is  independent of the selection schemes (SS or OS). 

In the following section, we present the SOP analysis in the backhaul KU scenario.

\section{SOP: Backhaul KU Scenario}\label{section_ku_sop}
This section presents the closed-form SOP and its asymptotic limit for both the SS and OS schemes in the backhaul KU scenario. The backhaul KU scenario is the one in which the knowledge of transmitters with active backhaul links is unavailable. Consequently, it is not known (in both SS and OS schemes) whether the backhaul link corresponding to the selected transmitter is active or not. Hence, in contrast to the KA scenario, the backhaul reliability factor is taken into account only after selecting a transmitter.  Thus, we define
$\hat{\Gamma}_{S}^{(k^*)} = \Gamma_{S}^{(k^*)} \mathbb{I}^{(k^*)}$.
We note that $\hat{\Gamma}_{S}^{(k^*)}$ follows a mixture distribution of $\Gamma_{S}^{(k^*)}$ and $\mathbb{I}^{(k^*)}$ similar to $\hat{\gamma}_{D}^{(k^*)}$ in Section \ref{section_sop_ka_scenario}. Therefore, the distribution is expressed as
\begin{align}\label{eq_CDF_Gamma_S_BHKU}
    &F_{\hat{\Gamma}_{S}^{(k^*)}}(x)=1-\zeta +\zeta F_{\Gamma_{S}^{(k^*)}}(x).
\end{align} 
 Utilizing $F_{\hat{\Gamma}_{S}^{(k^*)}}(x)$ corresponding to the particular selection schemes, in the following sections, we derive the SOP in the backhaul KU scenario with the help of  \eqref{eq_SOP_deinition}.
 
\subsection{SOP of the SS scheme}\label{subsection_sop_ss_ku}
In the backhaul KU scenario, the SS scheme selects a transmitter for which the SNR at the destination is maximum without considering which backhauls are active. Thus, without the reliability factor, we define
\begin{align}\label{eq_Gamma_sk_SS_BHKU}
\Gamma_{S}^{(k^*)}=\frac{1+\max\limits_{k\in \mathcal{S}}\{\gamma_{D}^{(k)}\}}{1+\gamma_{E}^{(k^*)}},
\end{align}
where $\gamma_{E}^{(k^*)}$ follows the distribution as in \eqref{eq_cdf_mult_eaves}. With the help of \eqref{eq_CDF_Gamma_S_BHKU} and the definition of $\Gamma_{S}^{(k^*)}$ in \eqref{eq_Gamma_sk_SS_BHKU}, we evaluate  
 $F_{\hat{\Gamma}_{S}^{(k^*)}}(x)$ including the reliability factor as 
\begin{align}\label{eq_CDF_GAMMA_S_SS_BHKU}
F_{\hat{\Gamma}_{S}^{(k^*)}}(x)
&=1-\zeta\sum_{k=1}^K\sum_{n=0}^{N-1} \sum_{\lb(\mathbf{m_D},\mathbf{\mu},\mathbf{l},\mathbf{m_E}\rb)\in\mathcal{X}} (-1)^{k+1} \binom{K}{k}\nn\\
&\times \Upsilon\frac{x^{\widehat{m}_{D}^{(k)} -\widehat{l}^{(k)}}\exp\big(-\frac{k x}{\lambda_D}\big)}{\big(x+\frac{(n+1)\lambda_D}{k \lambda_E}\big)^{\theta^{(k,n)}}},
\end{align}
where $\Upsilon, \mathbf{m_D},\mathbf{\mu},\mathbf{l}$, and $\mathbf{m_E}$ are defined in \eqref{eq_Upsilon_SS_BHKA}-\eqref{eq_cross_mul_SS_BHKA_2nd_time}. The solution approach is similar to \eqref{eq_CDF_GAMMA_S_SS_BHKA_int_form}.
The closed-form SOP is obtained easily using \eqref{eq_CDF_GAMMA_S_SS_BHKU} with the help of  \eqref{eq_SOP_deinition}.
We note that \eqref{eq_CDF_GAMMA_S_SS_BHKU} is similar to \eqref{eq_CDF_GAMMA_S_SS_BHKA_sol}, except $\zeta$ is just a multiplier in the second term of the expression in \eqref{eq_CDF_GAMMA_S_SS_BHKU},  whereas  $\zeta$ is raised to the power $k$ in \eqref{eq_CDF_GAMMA_S_SS_BHKA_sol}.  How this difference affects the SOP in the KU scenario, we investigate through the asymptotic analysis in the following section.

\subsection{Asymptotic analysis of the SS scheme }\label{subsection_asymp_sop_ss_ku}

With the assumption for the asymptotic analysis as in Section \ref{subsection_asymp_sop_ss_ka}, we obtain  $F_{{\gamma}_{D}^{(k)}}(x)=0$ from \eqref{eq_CDF_D}. Using this in \eqref{eq_CDF_Gamma_S_BHKU}, $F_{\Gamma_{S}^{(k^*)}}(x)$ becomes zero and thus the asymptotic SOP is obtained as 
\begin{align}\label{eq_SOP_Asympt_SS_BHKU}
  P_{out}^{\infty}= 1-\zeta.
\end{align}
From the above equation, we observe that the asymptotic SOP saturates to a constant value governed by $\zeta$. This is in contrast to the observation in the SS scheme for the KA scenario in \eqref{eq_SOP_Asympt_SS_BHKA}, 
where the asymptotic SOP is determined by both $\zeta$ and $K$.
 
\subsection{SOP of the OS scheme}\label{subsection_sop_os_ku}
To obtain the SOP as in Section \ref{subsection_sop_os_ka}, we need the CDF of $\hat{\Gamma}_{S}^{(k^*)}$ for the OS scheme in \eqref{eq_CDF_Gamma_S_BHKU}. The definition of $\Gamma_{S}^{(k^{*})}$ in the OS scheme without the backhaul reliability factor is 
\begin{align}\label{eq_Gamma_sk_OS_BHKU}
\Gamma_{S}^{(k^*)}= 
\max_{k\in \mathcal{S}}\bigg\{\frac{1+\gamma_{D}^{(k)}}{1+\gamma_{E}^{(k)}}\bigg\}.
\end{align}
 With the help of \eqref{eq_CDF_Gamma_S_BHKU} and the definition of $\Gamma_{S}^{(k^*)}$ in \eqref{eq_Gamma_sk_OS_BHKU}, we evaluate  
 \begin{align}\label{eq_CDF_GAMMA_S_OS_BHKU}
& F_{\hat{\Gamma}_{S}^{(k^*)}}(x) =1-\zeta+\zeta\bigg[1-\sum_{m_D =0}^{M_D-1}\sum_{\mu=0}^{m_D}\sum_{l=0}^{m_D-\mu}\sum_{n=0}^{N-1}\sum_{\mathbf{m_E}\in\mathcal{M}_{E}^{(n)}}\nn\\
 &\times (-1)^{n+l}\binom{m_D}{\mu}\binom{m_D-\mu}{l}\binom{N-1}{n}\frac{ N {(\lambda_D)}^{\varphi^{(n)}-m_D}}{(\lambda_E)^{M_E+\widehat{m}_{E}^{(n)}}}\nn\\
   &\times \frac{\Gamma(\varphi^{(n)})x^{m_D-l}\exp\big(-\frac{x-1}{\lambda_D}\big)}{\Gamma(M_E)(m_D!)\Big(\prod\limits_{i=1}^{n}m_{E}^{(i)}!\Big)\big(x+\frac{(n+1)\lambda_D}{\lambda_E}\big)^{\varphi^{(n)}}}\bigg]^K.
\end{align}
The closed-form SOP can be obtained using \eqref{eq_CDF_GAMMA_S_OS_BHKU} with the help of  \eqref{eq_SOP_deinition}.
 We conclude from \eqref{eq_CDF_GAMMA_S_OS_BHKU} that a fixed bias of $1-\zeta$ prevents the SOP from approaching zero as $K$ increases. The next section verifies this through the asymptotic analysis.  The same observation was made through the asymptotic analysis of the OS scheme in the KA scenario in Section \ref{subsection_sop_os_ka} where the SOP approached 
$1-\zeta$.

\subsection{Asymptotic analysis of the OS scheme}
\label{subsection_asymp_sop_os_ku}
Following the method utilized for the asymptotic analysis in Section \ref{subsection_asymp_sop_ss_ku}, it can be shown that $P_{out}^{\infty}=1-\zeta$ as in \eqref{eq_SOP_Asympt_SS_BHKU} for the SS scheme.
Here, we notice that irrespective of the selection scheme (SS or OS), the asymptotic SOP in the backhaul KU scenario is governed by $\zeta$ only. The observation is in contrast to that of the backhaul KA scenario, where it was shown that the asymptotic SOP depends on both $\zeta$ and $K$.  
We also conclude that the asymptotic SOP is unaffected by the number of eavesdroppers irrespective of the selection scheme (SS or OS) or availability of the backhaul activity knowledge (KA or KU).

In the subsequent sections, we evaluate the ESR.

\section{ESR: KA Scenario}\label{section_esr_ka}
In this section, we study the impact of the backhaul activity knowledge on the ESR in both selection schemes. The exact closed-form ESR analysis is followed by its asymptotic ESR analysis. 

\subsection{ESR of the SS scheme}\label{subsection_esr_ss_ka}
With the help of the 
CDF $F_{\hat{\Gamma}_{S}^{(k^*)}}(x)$ given by \eqref{eq_CDF_GAMMA_S_SS_BHKA_sol}, we evaluate the ESR in the SS scheme using \eqref{eq_ESR_CDF_eqn} as
\begin{align}\label{eq_C_erg_SS_BHKA_int_form}
 C_{erg} &=\frac{1}{\ln(2)} \sum_{k=1}^K\sum_{n=0}^{N-1} \sum_{\lb(\mathbf{m_D},\mathbf{\mu},\mathbf{l},\mathbf{m_E}\rb)\in\mathcal{X}}(-1)^{k+1} \binom{K}{k}\zeta^k \Upsilon\nn\\
 & \times 
  \underbrace{\int_{1}^{\infty} \frac{x^{\widehat{m}_{D}^{(k)} -\widehat{l}^{(k)}-1}\exp\big(-\frac{k x}{\lambda_D}\big)}{\left(x+\frac{(n+1)\lambda_D}{k \lambda_E}\right)^{\theta^{(k,n)}}}dx}_{\mathcal{I}},
\end{align}
where $\Upsilon, \mathbf{m_D},\mathbf{\mu},\mathbf{l}$, and $\mathbf{m_E}$ are defined in \eqref{eq_Upsilon_SS_BHKA}-\eqref{eq_cross_mul_SS_BHKA_2nd_time}. The solution of integral $\mathcal{I}$ in \eqref{eq_C_erg_SS_BHKA_int_form} depends on the  power of $x$ in the numerator. Accordingly, $\mathcal{I}$ in \eqref{eq_C_erg_SS_BHKA_int_form} is expressed as
\begin{align}\label{eq_I_I_0_I_1}    \mathcal{I}= \mathcal{I}_{0}^{(k)}+\mathcal{I}_{1}^{(k)}.
\end{align}
In \eqref{eq_I_I_0_I_1}, $\mathcal{I}_0^{(k)}$ corresponds to the condition $(\widehat{m}_{D}^{(k)} -\widehat{l}^{(k)})=0$ with its definition and respective solution given by
\begin{align}\label{eq_I_0_SS_BHKA}
\mathcal{I}_{0}^{(k)}&=\int_{1}^{\infty} \frac{\exp\big(-\frac{k x}{\lambda_D}\big)}{x\big(x+\frac{(n+1)\lambda_D}{k \lambda_E}\big)^{\theta^{(k,n)}}}dx\\
\label{eq_I_0_SS_BHKA_sol}
&=A \Gamma\left(0,\frac{k}{\lambda_D}\right)-\sum_{t=1}^{\theta^{(k,n)}}B^{(t)}\mathcal{W}_{(t-1)}.
\end{align}
To solve of integral in \eqref{eq_I_0_SS_BHKA} we used partial fractions of the denominator, and the integral solution \cite[eq. (3.462.16) and (3.462.19)]{ryzhik_2007}.
$\mathcal{I}_1^{(k)}$ in \eqref{eq_I_I_0_I_1} corresponds to the condition $(\widehat{m}_{D}^{(k)} -\widehat{l}^{(k)})>0$ with its definition and respective solution given by
\begin{align}\label{eq_I_1_SS_BHKA}
\mathcal{I}_{1}^{(k)}&=\int_{1}^{\infty} \frac{x^{\widehat{m}_{D}^{(k)}-\widehat{l}^{(k)}-1}\exp\left(-\frac{k x}{\lambda_D}\right)}{\left(x+\frac{(n+1)\lambda_D}{k \lambda_E}\right)^{\theta^{(k,n)}}}dx\\
\label{eq_I_1_SS_BHKA_sol}
&=\sum_{j=0}^{\widehat{m}_{D}^{(k)}-\widehat{l}^{(k)})-1}C^{(j)}\mathcal{W}_{(\theta^{(k,n)}-j-1)}.
\end{align}
To solve $\mathcal{I}_{1}^{(k)}$ in \eqref{eq_I_1_SS_BHKA}, we need to consider that the power of $x$ in the numerator of \eqref{eq_I_1_SS_BHKA} may be greater than the highest power in the denominator. Therefore, we employ a change of variable, binomial expansion, and the use of integral solution \cite[eq. (3.351.2)]{ryzhik_2007}. 
In \eqref{eq_I_0_SS_BHKA_sol} and \eqref{eq_I_1_SS_BHKA_sol}, $\mathcal{W}_{(\Theta)} \triangleq \left(\frac{k}{\lambda_D}\right)^{\Theta}\exp{\left(\frac{(n+1)}{\lambda_E}\right)}\Gamma\left(-\Theta,\frac{k}{\lambda_D}\left(1+\frac{(n+1)\lambda_D}{k\lambda_E}\right)\right)$, and the partial fraction coefficients $A = \left(\frac{k \lambda_E}{(n+1)\lambda_D}\right)^{\theta^{(k,n)}}$, $B^{(t)}= \left(\frac{k \lambda_E}{(n+1)\lambda_D}\right)^{\theta^{(k,n)}-t+1}$, and  $C^{(j)}=\binom{\widehat{m}_{D}^{(k)}-\widehat{l}^{(k)}-1}{j}\left(-\frac{(n+1)\lambda_D}{k \lambda_E}\right)^{\widehat{m}_{D}^{(k)}-\widehat{l}^{(k)}-j-1}$.

The ESR expression in \eqref{eq_C_erg_SS_BHKA_int_form} is more computationally intensive than the SOP expression in \eqref{eq_CDF_GAMMA_S_SS_BHKA_sol}, mainly since \eqref{eq_C_erg_SS_BHKA_int_form} involves the incomplete Gamma functions. It is also difficult to understand how the ESR depends on system parameters $K$, $N$, $M_D$, $M_E$, $\lambda_D$, and $\lambda_E$. Therefore, we present the asymptotic analysis of the ESR in the following section.
\subsection{Asymptotic Analysis of the SS Scheme}\label{subsection_hsnr_esr_ss_ka}
For the asymptotic analysis, we derive a simplified high-SNR ESR expression first. For that, we assume $\gamma_D^{(k)},\gamma_E^{(k)}>>1$ to obtain  $F_{\hat{\Gamma}_{S}^{(k^*)}}(x)$. Next, by assuming $\lambda_D>>\lambda_E$, we obtain the asymptotic ESR, $C_{erg}^\infty$ . With the assumption $\gamma_D^{(k)},\gamma_E^{(k)}>>1$, we remove unity from the numerator and the denominator of \eqref{eq_Gamma_sk_SS_BHKA} and use \eqref{eq_CDF_GAMMA_S_SS_BHKA_int_form} to obtain  $F_{\hat{\Gamma}_{S}^{(k^*)}}(x)$ as

\begin{align}\label{eq_Gamma_S_BHKA_SS_High_SNR}
F_{\hat{\Gamma}_{S}^{(k^*)}}(x)
&=1-\sum_{k=1}^{K}\sum_{n=0}^{N-1}\sum_{\lb(\mathbf{m_D},\mathbf{m_E}\rb)\in\mathcal{X}}(-1)^{k+1}\binom{K}{k}\zeta^k\Upsilon\nn\\
&\times \frac{x^{\widehat{m}_{D}^{(k)}}}{\big(x+\frac{(n+1)\lambda_D}{k \lambda_E}\big)^{\phi^{(k,n)}}},
\end{align}
where
\begin{align}
 \label{eq_Upsilon_HSNR_SS_BKHA}
\Upsilon& \triangleq  \frac{(-1)^{n}\binom{N-1}{n}N{\big(\frac{\lambda_D}{\lambda_E}\big)}^{M_E+\widehat{m}_{E}^{(n)}}\Gamma \left(\phi^{(k,n)}\right)}{k^{\phi^{(k,n)}}\Gamma(M_E)\Big(\prod\limits_{q=1}^{k}m_{D}^{(q)}!\Big)\Big(\prod\limits_{i=1}^{n}m_{E}^{(i)}!\Big)},
\end{align}
and $\phi^{(k,n)}\triangleq M_E+\widehat{m}_{D}^{(k)}+\widehat{m}_{E}^{(n)}$, and $\mathbf{m_D}$, and $\mathbf{m_E}$ are defined in \eqref{eq_Upsilon_SS_BHKA}. Similar to \eqref{eq_CDF_GAMMA_S_SS_BHKA_sol}, we denote the set of vector tuples $(\mathbf{m_D},\mathbf{m_E})$ by $\mathcal{X}$ in \eqref{eq_Gamma_S_BHKA_SS_High_SNR}. With the substitution of \eqref{eq_Gamma_S_BHKA_SS_High_SNR} into \eqref{eq_ESR_CDF_eqn}, the high-SNR ESR is evaluated as
\begin{align}\label{eq_C_erg_SS_int_form}
C_{erg} 
&=\frac{1}{\ln(2)} \sum_{k=1}^K\sum_{n=0}^{N-1}\sum_{\lb(\mathbf{m_D},\mathbf{m_E}\rb)\in\mathcal{X}} (-1)^{k+1} \binom{K}{k} \zeta^k\Upsilon\nn\\
&\times \big[\mathcal{I}_{0}^{(k)}+\mathcal{I}_{1}^{(k)}\big].
\end{align}
Similar to  \eqref{eq_C_erg_SS_BHKA_int_form}, $\mathcal{I}_0^{(k)}$ in \eqref{eq_C_erg_SS_int_form} corresponds to the integral with condition $\widehat{m}_{D}^{(k)}=0$, with the definition and the corresponding solution given by
\begin{align}\label{eq_I_0_HSNR_SS_BHKA_int_form}
&\mathcal{I}_{0}^{(k)}=\int_{1}^{\infty} \frac{1}{x\big(x+\frac{(n+1)\lambda_D}{k \lambda_E}\big)^{\phi^{(k,n)}}}dx,\\
\label{eq_I_0_High_SNR_SS_BHKA}
&
= A \ln\Big(1+\frac{(n+1)\lambda_D}{k \lambda_E}\Big) \nn\\
&-\sum_{t=2}^{\phi^{(k,n)}} \frac{B^{(t)}}{(t-1)\big(1+\frac{(n+1)\lambda_D}{k \lambda_E}\big)^{t-1}}.
\end{align}
In \eqref{eq_I_0_High_SNR_SS_BHKA}, the solution approach is similar to that adopted in  \eqref{eq_I_0_SS_BHKA} with the partial fraction coefficients $A= \left(\frac{k \lambda_E}{(n+1)\lambda_D}\right)^{\phi^{(k,n)}}$, and $B^{(t)}= \left(\frac{k \lambda_E}{(n+1)\lambda_D}\right)^{\phi^{(k,n)}-t+1}$.
Also, as in  \eqref{eq_C_erg_SS_BHKA_int_form}, $\mathcal{I}_1^{(k)}$ in \eqref{eq_C_erg_SS_int_form} corresponds to the integral with condition $\widehat{m}_{D}^{(k)}\geq 1$ with the definition and its corresponding solution given by
\begin{align}\label{eq_I_1_HSNR_SS_BHKA_int_form}
&\mathcal{I}_{1}^{(k)}=\int_{1}^{\infty} \frac{x^{\widehat{m}_{D}^{(k)}-1}}{\big(x+\frac{(n+1)\lambda_D}{k \lambda_E}\big)^{\phi^{(k,n)}}}dx\\
\label{eq_I_1_High_SNR_SS_BHKA}
&=\sum_{j=0}^{\widehat{m}_{D}^{(k)}-1}  \frac{C^{(j)}}{\left(\phi^{(k,n)}-j-1\right)\big(1+\frac{(n+1)\lambda_D}{k \lambda_E}\big)^{M_E+\widehat{m}_{E}^{(n)}-j-1}}.
\end{align}
The solution of $\mathcal{I}_{1}^{(k)}$ in  \eqref{eq_I_1_High_SNR_SS_BHKA} is obtained using a change of variable as the  highest power of $x$ in the numerator of \eqref{eq_I_1_HSNR_SS_BHKA_int_form} is always smaller than any power of $x$ in the denominator. $C^{(j)}=(-1)^{\widehat{m}_{D}^{(k)}-j-1}\binom{\widehat{m}_{D}^{(k)}-1}{j}\left(\frac{(n+1)\lambda_D}{k \lambda_E}\right)^{\widehat{m}_{D}^{(k)}-j-1}$ is the partial fraction coefficient specific to \eqref{eq_I_1_High_SNR_SS_BHKA}.

The absence of incomplete Gamma function and two interdependent summations in the solution of 
$\mathcal{I}_0^{(k)}$ and $\mathcal{I}_1^{(k)}$ in \eqref{eq_I_0_High_SNR_SS_BHKA} and \eqref{eq_I_1_High_SNR_SS_BHKA} compared to \eqref{eq_I_0_SS_BHKA}-\eqref{eq_I_1_SS_BHKA} make the high-SNR ESR computation far less complex than the exact ESR. Also from \eqref{eq_C_erg_SS_int_form}, we notice that the ESR is a weighted sum of the ratio $\lambda_D/\lambda_E$, which otherwise is difficult to understand from \eqref{eq_C_erg_SS_BHKA_int_form}. Note that the ratio $\lambda_D/\lambda_E$ quantifies the relative channel qualities of the $\text{S}^{(k)}$-$\text{D}$ and $\text{S}^{(k)}$-$\text{E}^{(n)}$ links. 


Next, we proceed to determine $C_{erg}^\infty$ by assuming $\lambda_D>>\lambda_E$ in \eqref{eq_C_erg_SS_int_form}. This assumption leads to a tight approximation $1+\frac{(n+1)\lambda_{D}}{k\lambda_{E}} \approx \frac{(n+1)\lambda_{D}}{k\lambda_{E}}$ in both $\mathcal{I}_0^{(k)}$ and $\mathcal{I}_1^{(k)}$ in 
\eqref{eq_I_0_High_SNR_SS_BHKA} and \eqref{eq_I_1_High_SNR_SS_BHKA}, respectively. The simplified $\mathcal{I}_{0}^{(k)}$ and $\mathcal{I}_{1}^{(k)}$ are then substituted into \eqref{eq_C_erg_SS_int_form}, leading to   $C_{erg}^{\infty}$ in \eqref{eq_simplifed_ESR_Asymp_SS_BHKA}. 
\begin{table*}
\begin{align}\label{eq_simplifed_ESR_Asymp_SS_BHKA}
  C_{erg}^{\infty} &=\frac{1}{\ln(2)} \sum_{k=1}^K \sum_{n=0}^{N-1}\sum_{\lb(\mathbf{m_D},\mathbf{m_E}\rb)\in\mathcal{X}}(-1)^{k+n+1}\binom{K}{k}\binom{N-1}{n} \frac{\zeta^k N\Gamma \big(\phi^{(k,n)}\big)}{{(k)}^{\widehat{m}_{D}^{(k)}}(n+1)^{M_E+\widehat{m}_{E}^{(n)}}\Gamma(M_E)}\nn\\
         &\times \frac{1}{\Big(\prod\limits_{q=1}^{k}m_{D}^{(q)}!\Big)\Big(\prod\limits_{i=1}^{n}m_{E}^{(i)}!\Big)}\Bigg[\underbrace{\ln\left(\frac{(n+1)\lambda_D}{k \lambda_E}\right)
 -H_{M_E+\widehat{m}_{E}^{(n)}-1}}_{\text{for}~ (\widehat{m}_{D}^{(k)})=0}+\underbrace{\sum_{j=0}^{\widehat{m}_{D}^{(k)}-1}  \frac{(-1)^{\widehat{m}_{D}^{(k)}-j-1\binom{\widehat{m}_{D}^{(k)}-1}{j}}}{\left(\phi^{(k,n)}-j-1\right)}}_{\text{for}~ (\widehat{m}_{D}^{(k)})\geq 1}\Bigg].
\end{align}
\hrule
\end{table*}
In \eqref{eq_simplifed_ESR_Asymp_SS_BHKA},  $H_{M_E+\widehat{m}_{E}^{(n)}-1}$ is the $(M_E+\widehat{m}_{E}^{(n)}-1)$-th harmonic number.
In the Section \ref{subsection_asymp_sop_ss_ka}, we observed that $P_{out}^\infty$ is solely limited by  $\zeta$ and $K$. Whereas from \eqref{eq_simplifed_ESR_Asymp_SS_BHKA}, we observe that $C_{erg}^\infty$  is not limited  by these factors. 
\subsection{ESR of the OS scheme}\label{subsection_esr_os_ka}
Here, we derive the ESR in the OS scheme by using the definition of $\hat{\Gamma}_{S}^{(k^*)}$ as in \eqref{eq_Gamma_sk_os_ka}. 
To evaluate the ESR using \eqref{eq_ESR_CDF_eqn}, we convert $F_{\hat{\Gamma}_{S}^{(k^*)}}(x)$ derived in \eqref{eq_CDF_GAMMA_S_OS_BHKA_sol} into a suitable form as
\begin{align}\label{eq_CDF_Gamma_S_OS_BHKA_psi_form}
     F_{\hat{\Gamma}_{S}^{(k^*)}}(x)&=1-\sum _{k=1}^K \sum_{\lb(\mathbf{m_D},\mathbf{\mu},\mathbf{l},\mathbf{n},\mathbf{m_E}\rb)\in\mathcal{X}}(-1)^{k+1} \binom{K}{k} \zeta^k \nn\\
     &\times \Upsilon\frac{x^{\widehat{m}_{D}^{(k)} -\widehat{l}^{(k)}}\exp\big(-\frac{kx}{\lambda_D}\big)}{\prod\limits_{q=1}^{k}\left(x+\frac{(n^{(q)}+1)\lambda_D}{\lambda_E}\right)^{\alpha^{(q)}}},
\end{align}
where
\begin{align}\label{eq_upsilon_OS}
&\Upsilon\triangleq \prod\limits_{q=1}^k (-1)^{n^{(q)}+l^{(q)}} \binom{m_{D}^{(q)}}{\mu^{(q)}}\binom{m_{D}^{(q)}-\mu^{(q)}}{l^{(q)}}\binom{N-1}{n^{(q)}}
\nn\\
  &\times
  \frac{ N{(\lambda_D)}^{\alpha^{(q)}-m_{D}^{(q)}}\Gamma\big(\alpha^{(q)}\big)\exp\big(\frac{1}{\lambda_D}\big)}{(m_{D}^{(q)}!)(\lambda_E)^{M_E+\sum_{i=1}^{n^{(q)}}m_{E}^{(q,i)}}\Big(\prod\limits_{i=1}^{n^{(q)}}m_{E}^{(q,i)}!\Big)\Gamma(M_E)},
\end{align}
and $\alpha^{(q)}=M_E+\mu^{(q)}+\sum_{i=1}^{n^{(q)}}{m_{E}^{(q,i)}}$. 
The results in \eqref{eq_CDF_Gamma_S_OS_BHKA_psi_form} are obtained with the help of the binomial expansion of $F_{\hat{\Gamma}_{S}}(x)$ in \eqref{eq_CDF_GAMMA_S_OS_BHKA_sol} and then converting the product-of-sums into a sum-of-products as
\begin{align}\label{eq_cross_mul_OS_BHKA_2nd_time}
&\Big(\sum_{i=0}^{\tau}\sum_{j=0}^i \sum_{u=0}^{i-j}\sum_{v=0}^{V}\sum_{\mathbf{m_E}\in \mathcal{M}_{E}^{(n)}} f(i,j,u,v,w_{E}^{(p)})\Big)^{\kappa}\nn\\
&=\sum_{\mathbf{m_D} \in \mathcal{M}_{D}^{(\kappa)}} \sum_{\mathbf{\mu} \in \mathcal{U}^{(\kappa)}(\mathbf{m_D})}\sum_{\mathbf{l}\in\mathcal{L^{(\kappa)}(\mathbf{m_D,\mu})}}\sum_{\mathbf{n}\in\mathcal{N}^{(\kappa)}}\sum_{\mathbf{m_E} \in \mathcal{M}_{E}^{(\kappa,\mathbf{n})}}
\nn\\
& 
\Big(\prod\limits_{q=1}^{{\kappa}}f(m_{D}^{(q)},\mu^{(q)},l^{(q)},n^{(q)},m_{E}^{(q,n)})\Big), 
\end{align}
where 
$\mathcal{M}_{D}^{(\kappa)}$, $\mathcal{U}^{(\kappa)}(\mathbf{m_D})$, and $\mathcal{L}^{(\kappa)}(\mathbf{m_D,\mu})$ have the definition same as in \eqref{eq_cross_mul_SS_BHKA_2nd_time}, $\mathcal{N}^{(\kappa)}$ is a set of integer vectors $[n^{(1)},\ldots,n^{(\kappa)}]$ containing $\kappa$ elements such that  $n^{(q)}\in\{0,\ldots,N-1\}$, $\mathcal{M}_{E}^{(\kappa,\mathbf{n})}$ is a set of index vectors $[m_{E}^{(1,1)} ,\ldots,  m_E^{(\kappa,n_i)}]$ containing $\kappa \times n_i$ elements such that $m_{E}^{(q,p)} \in \{0 ,\ldots, (M_E-1)\}$ for each $q\in\{1 ,\ldots, \kappa\}$,  and $p\in\{1 ,\ldots, n_i\}$ for each $i\in\{1 ,\ldots, \kappa\}$. Also, in \eqref{eq_CDF_Gamma_S_OS_BHKA_psi_form} we denote by $\mathcal{X}$ the set of vector tuples $\lb(\mathbf{m_D},\mathbf{\mu},\mathbf{l},\mathbf{n},\mathbf{m_E}\rb)$. Substituting \eqref{eq_CDF_Gamma_S_OS_BHKA_psi_form} in \eqref{eq_ESR_CDF_eqn}, the ESR is evaluated as
\begin{align}\label{eq_C_erg_OS_BHKA_int_form}
C_{erg} &
=\frac{1}{\ln(2)} \sum_{k=1}^K \sum_{\lb(\mathbf{m_D},\mathbf{\mu},\mathbf{l},\mathbf{n},\mathbf{m_E}\rb)\in\mathcal{X}} (-1)^{k+1} \binom{K}{k}\zeta^k \Upsilon\nn\\
&\times \big[\mathcal{I}_0^{(k)}+\mathcal{I}_1^{(k)}\big]
. 
\end{align}
Similar to \eqref{eq_C_erg_SS_BHKA_int_form}, $\mathcal{I}_0^{(k)}$ and $\mathcal{I}_1^{(k)}$ correspond to the integrals with the conditions $(\widehat{m}_{D}^{(k)} -\widehat{l}^{(k)})=0$ and $(\widehat{m}_{D}^{(k)} -\widehat{l}^{(k)})\geq1$, respectively, and are defined as 
\begin{align}\label{eq_I_0_OS_BHKA_int_form}
\mathcal{I}_{0}^{(k)}&=\int_1^\infty \frac{\exp\big(-\frac{kx}{\lambda_D}\big)}{x\Big(\prod\limits_{q=1}^{k}\big(x+\frac{(n^{(q)}+1)\lambda_D}{\lambda_E}\big)^{\alpha^{(q)}}\Big)}dx,\\
\label{eq_I_1_OS_BHKA}
\mathcal{I}_{1}^{(k)}&=\int_1^\infty \frac{x^{\widehat{m}_{D}^{(k)} -\widehat{l}^{(k)}-1}\exp\big(-\frac{kx}{\lambda_D}\big)}{\prod\limits_{q=1}^{k}\left(x+\frac{(n^{(q)}+1)\lambda_D}{\lambda_E}\right)^{\alpha^{(q)}}}dx.
\end{align}

To solve the integral $\mathcal{I}_{0}^{(k)}$,  we notice that the values of $n^{(q)}$ in the denominator of \eqref{eq_I_0_OS_BHKA_int_form} may be the same or different for the distinct values of  $q\in\{1 ,\ldots,  k\}$. We  use  $\mathcal{Z}$ to denote the number of distinct values of $n^{(q)}$ that are taken on more than once. We define $\mathcal{Q}_i$, where $i\in\{1 ,\ldots, \mathcal{Z}\}$, as the set of indices $q$ for which $n^{(q)}$ takes the $i^{\text{th}}$ distinct value. We also define $\mathcal{Q}=\mathcal{Q}_1\cup\mathcal{Q}_2\cup ,\ldots, \mathcal{Q}_\mathcal{Z}$ and $\bar{\mathcal{Q}}=\{1 ,\ldots, k\}-\mathcal{Q}$. Note that when the values of $n^{(q)}$ for each $q\in\{1,\ldots,k\}$ are distinct, $\bar{\mathcal{Q}}$ is an empty set. With these notations, we find the partial fraction expansions of the integrand, and with the help of the integral solution, \cite[eq. (3.462.16) and (3.462.19)]{ryzhik_2007}, obtain the solution in \eqref{eq_I_0_OS_BHKA_sol}. The second term in 
\eqref{eq_I_0_OS_BHKA_sol} corresponds to the case of the values $n^{(q)}$ that appear more than once, while the third term corresponds to the case of the values of $n^{(q)}$ that appear exactly once. 

\begin{table*}[]
\begin{align}\label{eq_I_0_OS_BHKA_sol}
\mathcal{I}_{0}^{(k)}&=A\Gamma\big(0,\frac{k}{\lambda_D}\big)
+\sum_{i=1}^{\mathcal{Z}}\sum_{t=1|q\in\mathcal{Q}_i}^{\widehat{\alpha}^{(|\mathcal{Q}_i|)}}B^{(i,t)}\mathcal{T}_{(t-1)}+\sum_{q\in\bar{\mathcal{Q}}}\sum_{t=1}^{\alpha^{(q)}}C^{(q,t)}\mathcal{T}_{(t-1)},\\
\label{eq_I_1_OS_BHKA_all_n_equal_sol}
\mathcal{I}_{1}^{(k)}&=\sum_{j=0}^{\widehat{m}_{D}^{(k)} -\widehat{l}^{(k)}-1}\binom{\widehat{m}_{D}^{(k)} -\widehat{l}^{(k)}-1}{j}\Big(-\frac{(n^{(q)}+1)\lambda_D}{\lambda_E}\Big)^{\widehat{m}_{D}^{(k)} -\widehat{l}^{(k)}-1-j}\mathcal{T}_{(\widehat{\alpha}^{(k)}-j-1)},\\
\label{eq_I_1_OS_BHKA_all_n_not_equal_sol}
\mathcal{I}_{1}^{(k)}&=\sum_{i=1}^{\mathcal{Z}}\sum_{t=1|q\in\mathcal{Q}_i}^{\widehat{\alpha}^{(|\mathcal{Q}_i|)}}\sum_{j=0}^{\widehat{m}_{D}^{(k)} -\widehat{l}^{(k)}-1}\binom{\widehat{m}_{D}^{(k)} -\widehat{l}^{(k)}-1}{j}
    \Big(-\frac{(n^{(q)}+1)\lambda_D}{\lambda_E}\Big)^{\widehat{m}_{D}^{(k)} -\widehat{l}^{(k)}-1-j}B^{(i,t)}\mathcal{T}_{(t-j-1)}\nn\\
    &+\sum_{q\in\bar{\mathcal{Q}}}^{\mathcal{Z}}\sum_{t=1}^{\alpha^{(q)}}\sum_{j=0}^{\widehat{m}_{D}^{(k)} -\widehat{l}^{(k)}-1}\binom{\widehat{m}_{D}^{(k)} -\widehat{l}^{(k)}-1}{j}   
    \Big(-\frac{(n^{(q)}+1)\lambda_D}{\lambda_E}\Big)^{\widehat{m}_{D}^{(k)} -\widehat{l}^{(k)}-1-j} C^{(q,t)}\mathcal{T}_{(t-j-1)}.
\end{align}
    \hrule
\end{table*}
For the solution of $\mathcal{I}_{1}^{(k)}$  in \eqref{eq_I_1_OS_BHKA}, we need to consider that the power of $x$ in the numerator of \eqref{eq_I_1_OS_BHKA} may be greater than the highest power in the denominator. Also the values of $n^{(q)}$ for distinct $q\in\{1,\ldots,k\}$ may be same or distinct. Here, we adopt the solution method by first checking whether all $n^{(q)}$ are equal to each other in \eqref{eq_I_1_OS_BHKA}. If they are equal, we employ a change of variable and then use the integral solution \cite[eq. (3.351.2)]{ryzhik_2007} to obtain the solution in \eqref{eq_I_1_OS_BHKA_all_n_equal_sol}. 
If any of the $n^{(q)}$ for distinct $q\in\{1,\ldots,k\}$ is also distinct, we employ the partial fraction method without including the numerator, followed by the change of variable as was done for the solution of $\mathcal{I}_{0}^{(k)}$. Finally, we use the integration solution \cite[eq. (3.351.2)]{ryzhik_2007} to obtain the solution in \eqref{eq_I_1_OS_BHKA_all_n_not_equal_sol}.
In \eqref{eq_I_0_OS_BHKA_sol}-\eqref{eq_I_1_OS_BHKA_all_n_not_equal_sol}, $\mathcal{T}_{(\Theta)}\triangleq  \left(\frac{k}{\lambda_D}\right)^{\Theta}\exp\left(\frac{k(n^{(q)}+1)}{\lambda_E}\right) \Gamma\left(-\Theta,\frac{k}{\lambda_D}\left(1+\frac{(n^{(q)}+1)\lambda_D}{\lambda_E}\right)\right)$, $\widehat{\alpha}^{(|\mathcal{Q}_i|)}\triangleq\sum_{q\in\mathcal{Q}_i}{\alpha}^{(q)}$, $\widehat{\alpha}^{(k)}\triangleq\sum_{q=1}^{k}{\alpha}^{(q)}$, and the partial fraction coefficients $A$, $B^{(i,t)}$, and $C^{(q,t)}$ are specific to $\mathcal{I}_{0}^{(k)}$, and $\mathcal{I}_{1}^{(k)}$, which can be obtained easily for a given $K$, $N$, $M_D$, and $M_E$.

In \eqref{eq_C_erg_SS_BHKA_int_form} as well as \eqref{eq_C_erg_OS_BHKA_int_form}, we observe that the ESR is a weighted sum of $\zeta^k$ for $k=1, \ldots, K$. As $\zeta<1$, increasing $K$ is not equally beneficial to improve ESR in the backhaul KA scenario as in the case of active backhaul links where $\zeta=1$. The ESR in \eqref{eq_C_erg_OS_BHKA_int_form} as in  \eqref{eq_C_erg_SS_BHKA_int_form} contains several interdependent summations, multiplications, and incomplete Gamma functions making it computationally intensive. To reduce the computation and to better understand the ESR dependency on the system parameters such as $K$, $N$, $M_D$, $M_E$, $\lambda_D$, and $\lambda_E$, we present the asymptotic analysis of the ESR in the following section.
\subsection{Asymptotic Analysis of the OS Scheme}\label{subsection_hsnr_esr_os_ka}
For the asymptotic analysis of the OS scheme, we use the same assumptions as in Section \ref{subsection_hsnr_esr_ss_ka}.  After removing unity from the numerator and the denominator of \eqref{eq_Gamma_sk_os_ka} and using an approach similar to \eqref{eq_CDF_GAMMA_S_OS_BHKA} and  \eqref{eq_CDF_Gamma_S_OS_BHKA_psi_form} we obtain the simplified $F_{\hat{\Gamma}_{S}}(x)$ as
\begin{align}\label{eq_Gamma_S_BHKA_OS_High_SNR}
    F_{\hat{\Gamma}_{S}}(x)
   &=1-\sum _{k=1}^K  \sum_{\lb(\mathbf{m_D},\mathbf{n},\mathbf{m_E}\rb)\in\mathcal{X}} (-1)^{k+1} \binom{K}{k} \zeta^k \Upsilon\nn\\
   &\times \frac{x^{\widehat{m}_{D}^{(k)}}}{\left(x+\frac{(n^{(q)}+1)\lambda_D}{\lambda_E}\right)^{\beta^{(q)}}},
\end{align}
where
\begin{align}\label{eq_upsilon_OSH}  
\Upsilon&\triangleq\prod\limits_{q=1}^{k}
\frac{ (-1)^{n^{(q)}}\binom{N-1}{n^{(q)}}N\big(\frac{\lambda_D}{\lambda_E}\big)^{M_E+\sum_{i=1}^{n^{(q)}}m_{E}^{(q,i)}} \Gamma\big(\beta^{(q)}\big)}{(m_{D}^{(q)}!)\Big(\prod\limits_{i=1}^{n^{(q)}}m_{E}^{(q,i)}!\Big)\Gamma(M_E)},
\end{align}
and $\beta^{(q)}=M_E+{m}_{D}^{(q)}+\sum_{i=1}^{n^{(q)}}{m_{E}^{(q,i)}}$.  In \eqref{eq_Gamma_S_BHKA_OS_High_SNR}, $\mathcal{X}$ denotes the set of vector tuples $\lb(\mathbf{m_D},\mathbf{n},\mathbf{m_E}\rb)$, where the definitions of $\mathbf{m_D},\mathbf{n}$, and $\mathbf{m_E}$ are the same as in \eqref{eq_CDF_Gamma_S_OS_BHKA_psi_form}. 
By substituting \eqref{eq_Gamma_S_BHKA_OS_High_SNR} into \eqref{eq_ESR_CDF_eqn}, the corresponding high-SNR ESR is evaluated as
\begin{align}\label{eq_C_erg_OS_BHKA_High_SNR_int_form}
     C_{erg}   &=\frac{1}{\ln(2)} \sum_{k=1}^K \sum_{\lb(\mathbf{m_D},\mathbf{n},\mathbf{m_E}\rb)\in\mathcal{X}} (-1)^{k+1} \binom{K}{k} \zeta^k \Upsilon\nn\\
     & \times \big[\mathcal{I}_{0}^{(k)}+\mathcal{I}_{1}^{(k)}\big]. 
\end{align}
 Similar to  \eqref{eq_C_erg_SS_BHKA_int_form}, $\mathcal{I}_0^{(k)}$ and $\mathcal{I}_1^{(k)}$ in \eqref{eq_C_erg_OS_BHKA_High_SNR_int_form} correspond to the integrals with the conditions $\widehat{m}_{D}^{(k)}=0$ and $\widehat{m}_{D}^{(k)}\geq1$, respectively, and are defined as
\begin{align}\label{eq_I_0_High_SNR_OS_BHKA_int_form}
\mathcal{I}_{0}^{(k)}&=\int_1^\infty \frac{1}{x\Big(\prod\limits_{q=1}^{k}\big(x+\frac{(n^{(q)}+1)\lambda_D}{\lambda_E}\big)^{M_E+\sum_{i=1}^{n^{(q)}}{m_{E}^{(q,i)}}}\Big)}dx,\\
\label{eq_I_1_High_SNR_OS_BHKA_int_form}
\mathcal{I}_{1}^{(k)}&=\int_1^\infty \frac{x^{\widehat{m}_{D}^{(k)}-1}}{\prod\limits_{q=1}^{k}\big(x+\frac{(n^{(q)}+1)\lambda_D}{\lambda_E}\big)^{\beta^{(q)}}}dx.
\end{align}

\begin{table*}[]
\begin{align}\label{eq_I_0_High_SNR_OS_BHKA_sol}
\mathcal{I}_{0}^{(k)}&=\sum_{i=1}^{\mathcal{Z}}\Big(\sum_{t=1|q\in\mathcal{Q}_i}^{1}{B^{(i,t)}}\ln\big(1+\frac{(n^{(q)}+1)\lambda_D}{\lambda_E}\big)+\sum_{t=2|q\in\mathcal{Q}_i}^{|\mathcal{Q}_i| M_E+\sum_{q\in\mathcal{Q}_i}\sum_{i=1}^{n^{(q)}}{m_{E}^{(q,i)}}} \frac{B^{(i,t)}}{(t-1)\big(1+\frac{(n^{(q)}+1)\lambda_D}{\lambda_E}\big)^{t-1}}\Big)\nn\\
    &+\sum_{q\in\bar{\mathcal{Q}}}\Big(\sum_{t=1}^{1}{C^{(q,t)}}\ln\big(1+\frac{(n^{(q)}+1)\lambda_D}{\lambda_E}\big)+\sum_{t=2}^{M_E+\sum_{i=1}^{n^{(q)}}{m_{E}^{(q,i)}}} \frac{C^{(q,t)}}{(t-1)\big(1+\frac{(n^{(q)}+1)\lambda_D}{\lambda_E}\big)^{t-1}}\Big),\\
\label{eq_I_1_High_SNR_OS_BHKA_case_1_sol}
\mathcal{I}_{1}^{(k)}&=\sum_{j=0}^{\widehat{m}_{D}^{(k)}-1}\binom{\widehat{m}_{D}^{(k)}-1}{j}\frac{\Big(-\frac{(n^{(q)}+1)\lambda_D}{\lambda_E}\Big)^{\widehat{m}_{D}^{(k)}-1-j}}{(\widehat{\beta}^{(k)}-j-1)\big(1+\frac{(n^{(q)}+1)\lambda_D}{\lambda_E}\big)^{\widehat{\beta}^{(k)}-j-1}},\\
\label{eq_I_1_High_SNR_OS_BHKA_case_2_sol}
\mathcal{I}_{1}^{(k)}
&=\sum_{i=1}^{\mathcal{Z}}\Big(\sum_{t=1}^{1}B^{(i,t)}\ln\big(1+\frac{(n^{(q)}+1)\lambda_D}{\lambda_E}\big)+\sum_{t=2|q\in\mathcal{Q}_i}^{\widehat{\beta}^{(|\mathcal{Q}_i|)}}\frac{B^{(i,t)}}{(t-1)\big(1+\frac{(n^{(q)}+1)\lambda_D}{\lambda_E}\big)^{t-1}}\Big)\nn\\
    &+\sum_{q \in\bar{\mathcal{Q}}}\Big(\sum_{t=1}^{1}C^{(q,t)}\ln\big(1+\frac{(n^{(q)}+1)\lambda_D}{\lambda_E}\big)+\sum_{t=2}^{\beta^{(q)}}\frac{C^{(q,t)}}{(t-1)\big(1+\frac{(n^{(q)}+1)\lambda_D}{\lambda_E}\big)^{t-1}}\Big).
\end{align}
\hrule
\end{table*}
The solution of $\mathcal{I}_{0}^{(k)}$ is provided in \eqref{eq_I_0_High_SNR_OS_BHKA_sol}. The solution approach in \eqref{eq_I_0_High_SNR_OS_BHKA_sol} is similar to that of  \eqref{eq_I_0_OS_BHKA_sol} following the method of partial fractions.
When all $n^{(q)}$ being equal for all $q \in \{1,\ldots,k\}$ in \eqref{eq_I_1_High_SNR_OS_BHKA_int_form}, the power of $x$ in the denominator is always greater than that in the numerator. The integral in \eqref{eq_I_1_High_SNR_OS_BHKA_int_form} is evaluated after the change of variable with the solution provided in \eqref{eq_I_1_High_SNR_OS_BHKA_case_1_sol}. However, if for any $q\in\{1 ,\ldots, k\}$ in \eqref{eq_I_1_High_SNR_OS_BHKA_int_form}, $n^{(q)}$ is distinct, the integral in \eqref{eq_I_1_High_SNR_OS_BHKA_int_form} is evaluated adopting the method of partial fractions as in \eqref{eq_I_0_High_SNR_OS_BHKA_sol} with the solution provided in \eqref{eq_I_1_High_SNR_OS_BHKA_case_2_sol}.   $\widehat{\beta}^{(k)}\triangleq\sum_{q=1}^{k}\beta^{(q)}$, and $\widehat{\beta}^{(|\mathcal{Q}_i|)}\triangleq \sum_{q \in \mathcal{Q}_i}(M_E+m_{D}^{(q)}+\sum_{i=1}^{n^{(q)}}{m_{E}^{(q,i)}})$. $B^{(i,t)}$ and $C^{(q,t)}$ are the partial fraction coefficients specific to $\mathcal{I}_{0}^{(k)}$ and $\mathcal{I}_{1}^{(k)}$ in \eqref{eq_I_0_High_SNR_OS_BHKA_sol}-\eqref{eq_I_1_High_SNR_OS_BHKA_case_2_sol}, and can be obtained easily for given $K$, $N$, $M_D$, and $M_E$. Similar to that for the SS scheme as in \eqref{eq_C_erg_SS_int_form}, the high-SNR ESR in the OS scheme in \eqref{eq_C_erg_OS_BHKA_High_SNR_int_form} is a weighted sum of the ratio $\lambda_D/\lambda_E$. 

Next, as in Section \ref{subsection_hsnr_esr_ss_ka}, we use the approximation $1+\frac{(n^{(q)}+1)\lambda_{D}}{\lambda_{E}} \approx \frac{(n^{(q)}+1)\lambda_{D}}{\lambda_{E}}$ in \eqref{eq_I_0_High_SNR_OS_BHKA_sol}-\eqref{eq_I_1_High_SNR_OS_BHKA_case_2_sol} to further simplify $\mathcal{I}_{0}^{(k)}$ and $\mathcal{I}_{1}^{(k)}$ . In this section, $\mathcal{I}_{0}^{(k)}$ in \eqref{eq_I_0_High_SNR_OS_BHKA_sol} simplifies to
\begin{align}\label{eq_I_0_Asympt_OS_BHKA}
\mathcal{I}_{0}^{(k)}&=\sum_{i=1}^{\mathcal{Z}}\Big(\sum_{t=1|q\in\mathcal{Q}_i}^{1}{B^{(i,t)}}\ln\big(\frac{(n^{(q)}+1)\lambda_D}{\lambda_E}\big)\nn\\
&+\sum_{t=2|q\in\mathcal{Q}_i}^{|\mathcal{Q}_i| M_E+\sum_{q\in\mathcal{Q}_i}\sum_{i=1}^{n^{(q)}}{m_{E}^{(q,i)}}} \frac{B^{(i,t)}}{(t-1)\big(\frac{(n^{(q)}+1)\lambda_D}{\lambda_E}\big)^{t-1}}\Big)\nn\\
    &+\sum_{q\in\bar{\mathcal{Q}}}\Big(\sum_{t=1}^{1}{C^{(q,t)}}\ln\big(\frac{(n^{(q)}+1)\lambda_D}{\lambda_E}\big)\nn\\
    &+\sum_{t=2}^{M_E+\sum_{i=1}^{n^{(q)}}{m_{E}^{(q,i)}}} \frac{C^{(q,t)}}{(t-1)\big(\frac{(n^{(q)}+1)\lambda_D}{\lambda_E}\big)^{t-1}}\Big).
\end{align}
When all $n^{(q)}$ for all distinct $q \in \{1,\ldots,k\}$ are equal, $\mathcal{I}_{1}^{(k)}$ in \eqref{eq_I_1_High_SNR_OS_BHKA_case_1_sol} simplifies to
\begin{align}\label{eq_I_1_Asympt_OS_BHKA_case_1}
\mathcal{I}_{1}^{(k)}&=\sum_{j=0}^{\widehat{m}_{D}^{(k)}-1}(-1)^{\widehat{m}_{D}^{(k)}-1-j}\binom{\widehat{m}_{D}^{(k)}-1}{j}\nn\\
&\times \frac{\big(\frac{(n^{(q)}+1)\lambda_D}{\lambda_E}\big)^{\widehat{m}_{E}^{(k)}-\widehat{\beta}^{(k)}}}{(\widehat{\beta}^{(k)}-j-1)} ,
\end{align}
and when any of the $n^{(q)}$ is distinct for any $q\in\{1 ,\ldots, k\}$, $\mathcal{I}_{1}^{(k)}$  in \eqref{eq_I_1_High_SNR_OS_BHKA_case_2_sol} simplifies to
\begin{align}\label{eq_I_1_Asympt_OS_BHKA_case_2}
\mathcal{I}_{1}^{(k)}&=\sum_{i=1}^{\mathcal{Z}}\Big(\sum_{t=1}^{1}B^{(i,t)}\ln\big(\frac{(n^{(q)}+1)\lambda_D}{\lambda_E}\big)\nn\\
&+\sum_{t=2|q\in\mathcal{Q}_i}^{\widehat{\beta}^{(|\mathcal{Q}_i|)}}\frac{B^{(i,t)}}{(t-1)\big(\frac{(n^{(q)}+1)\lambda_D}{\lambda_E}\big)^{t-1}}\Big)\nn\\
    &+\sum_{q \in\mathcal{Q}}\Big(\sum_{t=1}^{1}C^{(q,t)}\ln\big(\frac{(n^{(q)}+1)\lambda_D}{\lambda_E}\big)\nn\\
    &+\sum_{t=2}^{\beta^{(q)}}\frac{C^{(q,t)}}{(t-1)\big(\frac{(n^{(q)}+1)\lambda_D}{\lambda_E}\big)^{t-1}}\Big).
\end{align}
Substituting \eqref{eq_I_0_Asympt_OS_BHKA} along with \eqref{eq_I_1_Asympt_OS_BHKA_case_1} and \eqref{eq_I_1_Asympt_OS_BHKA_case_2} in \eqref{eq_C_erg_OS_BHKA_High_SNR_int_form}, $C_{erg}^{\infty}$ is obtained.  
\section{ESR: Backhaul KU Scenario}\label{section_esr_ku}
In this section, we perform the ESR analysis in the backhaul KU scenario. Similar to Section \ref{section_esr_ka}, the derivation of the exact ESR is followed by its asymptotic ESR.
\subsection{ESR of the SS scheme}\label{subsection_esr_ss_ku}
Similar to the ESR analysis of the SS scheme for the KA scenario in Section \ref{subsection_esr_ss_ka}, we utilize  $F_{\hat{\Gamma}_{S}^{(k^*)}}(x)$ from \eqref{eq_CDF_GAMMA_S_SS_BHKU} in \eqref{eq_ESR_CDF_eqn} in the backhaul KU scenario.
 The ESR is then evaluated as
\begin{align}\label{eq_C_erg_SS_BHKU_int_form}
 C_{erg} &=\frac{1}{\ln(2)} \sum_{k=1}^K\sum_{n=0}^{N-1} \sum_{\lb(\mathbf{m_D},\mathbf{\mu},\mathbf{l},\mathbf{m_E}\rb)\in\mathcal{X}} (-1)^{k+1} \binom{K}{k}\zeta \Upsilon\nn\\
 &\times \big[\mathcal{I}_0^{(k)}+\mathcal{I}_1^{(k)}\big], 
\end{align}
where $\Upsilon, \mathbf{m_D},\mathbf{\mu},\mathbf{l}$, and $\mathbf{m_E}$ are defined in \eqref{eq_Upsilon_SS_BHKA}-\eqref{eq_cross_mul_SS_BHKA_2nd_time} and in \eqref{eq_C_erg_SS_BHKU_int_form}, $\mathcal{X}$ is used to denote the set of vector tuples $\lb(\mathbf{m_D},\mathbf{\mu},\mathbf{l},\mathbf{m_E}\rb)$. 
$\mathcal{I}_{0}^{(k)}$ in \eqref{eq_C_erg_SS_BHKU_int_form} is exactly the same as in \eqref{eq_I_0_SS_BHKA} and corresponds to the integral with condition $(\widehat{m}_{D}^{(k)} -\widehat{l}^{(k)})=0$ with its solution already provided in  \eqref{eq_I_0_SS_BHKA_sol}. Similarly  $\mathcal{I}_{1}^{(k)}$ in \eqref{eq_C_erg_SS_BHKU_int_form} corresponds to the condition $(\widehat{m}_{D}^{(k)} -\widehat{l}^{(k)})\geq1$ with the definition in  \eqref{eq_I_1_SS_BHKA} and its solution in \eqref{eq_I_1_SS_BHKA_sol}.  Note that the ESR in \eqref{eq_C_erg_SS_BHKU_int_form} is scaled down by a factor of $\zeta$ from all transmitters due to the unreliable backhaul KU scenario in contrast to the backhaul KA scenario in \eqref{eq_C_erg_SS_BHKA_int_form} where the  ESR is scaled down by $\zeta^k$ for $k=1,\ldots,K$ transmitters. 

\subsection{Asymptotic Analysis of the SS Scheme}\label{subsection_hsnr_esr_ss_ku}
After making the same assumptions as in Section \ref{subsection_hsnr_esr_ss_ka}, we first obtain a simplified expression for $F_{\hat{\Gamma}_{S}^{(k^*)}}(x)$ by removing unity from the numerator and the denominator of \eqref{eq_Gamma_sk_SS_BHKU}. Then, we proceed with steps similar to \eqref{eq_CDF_GAMMA_S_SS_BHKU} to obtain 
\begin{align}\label{eq_Gamma_S_BHKU_SS_HSNR}
    F_{\hat{\Gamma}_{S}^{(k^*)}}(x)
&=1-\sum_{k=1}^{K}\sum_{n=0}^{N-1}\sum_{\lb(\mathbf{m_D},\mathbf{m_E}\rb)\in\mathcal{X}}(-1)^{k+1}\zeta\binom{K}{k}\Upsilon\nn\\
&\times \frac{x^{\widehat{m}_{D}^{(k)}}}{\big(x+\frac{(n+1)\lambda_D}{k \lambda_E}\big)^{\phi^{(k,n)}}},
\end{align}
where $\Upsilon, \mathbf{m_D}$, and $\mathbf{m_E}$ are defined in \eqref{eq_Upsilon_SS_BHKA}-\eqref{eq_cross_mul_SS_BHKA_2nd_time} and in \eqref{eq_Gamma_S_BHKU_SS_HSNR}, $\mathcal{X}$ is used to denote the set of vector tuples $\lb(\mathbf{m_D},\mathbf{m_E}\rb)$. The corresponding high-SNR ESR is then evaluated by substituting \eqref{eq_Gamma_S_BHKU_SS_HSNR} into \eqref{eq_ESR_CDF_eqn} to obtain
\begin{align}\label{eq_C_erg_SS_BHKU_HSNR_int_form}
C_{erg} &=\frac{1}{\ln(2)} \sum_{k=1}^K\sum_{n=0}^{N-1}\sum_{\lb(\mathbf{m_D},\mathbf{m_E}\rb)\in\mathcal{X}} (-1)^{k+1} \binom{K}{k} \zeta\Upsilon\nn\\
& \times \big[\mathcal{I}_0^{(k)}+\mathcal{I}_1^{(k)}\big],
\end{align}
where the solutions of $\mathcal{I}_{0}^{(k)}$ and $\mathcal{I}_{1}^{(k)}$ are the same as in \eqref{eq_I_0_High_SNR_SS_BHKA} and \eqref{eq_I_1_High_SNR_SS_BHKA}, respectively, with their corresponding conditions.
For $C_{erg}^{\infty}$, we use the approximation  $1+\frac{(n+1)\lambda_{D}}{k\lambda_{E}} \approx \frac{(n+1)\lambda_{D}}{k\lambda_{E}}$ in 
 \eqref{eq_I_0_High_SNR_SS_BHKA} and \eqref{eq_I_1_High_SNR_SS_BHKA} to obtain the simplified $\mathcal{I}_{0}^{(k)}$ and $\mathcal{I}_{1}^{(k)}$, respectively. The resulting $C_{erg}^{\infty}$ is provided in \eqref{eq_simplifed_ESR_Asymp_SS_BHKU}.
\begin{table*}
\begin{align}\label{eq_simplifed_ESR_Asymp_SS_BHKU}
C_{erg}^{\infty}&=\frac{1}{\ln(2)} \sum_{k=1}^K \sum_{n=0}^{N-1}\sum_{\lb(\mathbf{m_D},\mathbf{m_E}\rb)\in\mathcal{X}}\binom{K}{k}\binom{N-1}{n} \frac{(-1)^{k+n+1}\zeta N\Gamma (\phi^{(k,n)})}{k^{\widehat{m}_{D}^{(k)}}(n+1)^{M_E+\widehat{m}_{E}^{(n)}}\Gamma(M_E)\Big(\prod\limits_{q=1}^{k}m_{D}^{(q)}!\Big)}\nn\\
         &\times \frac{1}{\Big(\prod\limits_{i=1}^{n}m_{E}^{(i)}!\Big)}\Big[\underbrace{\ln\Big(\frac{(n+1)\lambda_D}{k \lambda_E}\Big)
 -H_{M_E+\widehat{m}_{E}^{(n)}-1}}_{\text{for}~ (\widehat{m}_{D}^{(k)})=0}+\underbrace{\sum_{j=0}^{\widehat{m}_{D}^{(k)}-1}  \frac{(-1)^{\widehat{m}_{D}^{(k)}-j-1\binom{\widehat{m}_{D}^{(k)}-1}{j}}}{(\phi^{(k,n)}-j-1)}}_{\text{for}~ (\widehat{m}_{D}^{(k)})\geq 1}\Big].
\end{align}
\hrule
\end{table*}

\subsection{ESR of the OS scheme}\label{subsection_esr_os_ku}
For the ESR analysis of the OS scheme using \eqref{eq_ESR_CDF_eqn}, we utilize $F_{\hat{\Gamma}_{S}^{(k^*)}}(x)$ from \eqref{eq_CDF_GAMMA_S_OS_BHKU}.  Similar to \eqref{eq_CDF_Gamma_S_OS_BHKA_psi_form},  $F_{\hat{\Gamma}_{S}^{(k^*)}}(x)$ in \eqref{eq_CDF_GAMMA_S_OS_BHKU} is expressed as
\begin{align}\label{eq_CDF_Gamma_S_OS_BHKU_psi_form}
    F_{\hat{\Gamma}_{S}^{(k^*)}}(x)&=1-\sum _{k=1}^K \sum_{\lb(\mathbf{m_D},\mathbf{\mu},\mathbf{l},\mathbf{n},\mathbf{m_E}\rb)\in\mathcal{X}}(-1)^{k+1} \binom{K}{k} \zeta  \Upsilon \nn\\
    & \times \frac{x^{\widehat{m}_{D}^{(k)} -\widehat{l}^{(k)}}\exp\big(-\frac{kx}{\lambda_D}\big)}{\prod\limits_{q=1}^{k}\left(x+\frac{(n^{(q)}+1)\lambda_D}{\lambda_E}\right)^{\alpha^{(q)}}},
\end{align}
where $\Upsilon, \mathbf{m_D},\mathbf{\mu},\mathbf{l},\mathbf{n}$, and $\mathbf{m_E}$ are defined in \eqref{eq_upsilon_OS}-\eqref{eq_cross_mul_OS_BHKA_2nd_time}. In \eqref{eq_CDF_Gamma_S_OS_BHKU_psi_form}, $\mathcal{X}$ is used to denote the set of vector tuples $\lb(\mathbf{m_D},\mathbf{\mu},\mathbf{l},\mathbf{n},\mathbf{m_E}\rb)$.
With substitution of \eqref{eq_CDF_Gamma_S_OS_BHKU_psi_form} in \eqref{eq_ESR_CDF_eqn}, the ESR is then evaluated as
 \begin{align}\label{eq_ESR_BHKU_psi_form}
C_{erg} &=\frac{1}{\ln(2)} \sum_{k=1}^K \sum_{\lb(\mathbf{m_D},\mathbf{\mu},\mathbf{l},\mathbf{n},\mathbf{m_E}\rb)\in\mathcal{X}} (-1)^{k+1} \binom{K}{k} \zeta \Upsilon \nn\\
& \times \big[\mathcal{I}_0^{(k)}+\mathcal{I}_1^{(k)}\big],
\end{align}
where the solution of $\mathcal{I}_{0}^{(k)}$ is the same as in \eqref{eq_I_0_OS_BHKA_sol}, and the solution of $\mathcal{I}_{1}^{(k)}$ is the same as in \eqref{eq_I_1_OS_BHKA_all_n_equal_sol}-\eqref{eq_I_1_OS_BHKA_all_n_not_equal_sol}.
Similar to \eqref{eq_C_erg_SS_BHKU_int_form}, we observe that the ESR in \eqref{eq_ESR_BHKU_psi_form} is scaled down by $\zeta$ in contrast to the backhaul KA scenario in \eqref{eq_C_erg_OS_BHKA_int_form}, where the  ESR is scaled down by $\zeta^k$ for $k=1, \ldots, K$ transmitters. Thus, from Section \ref{section_esr_ka} and \ref{section_esr_ku}, we conclude that the influence of backhaul reliability on the ESR performance in the backhaul KU scenario is more than that in the backhaul KA scenario.


\subsection{Asymptotic Analysis of the OS Scheme}\label{subsection_hsnr_esr_os_ku}
For the asymptotic analysis of the ESR in the OS scheme as in Section \ref{subsection_hsnr_esr_os_ka}, we remove unity from the numerator and the denominator in the definition of $\Gamma_{S}^{(k^*)}$ in \eqref{eq_Gamma_sk_OS_BHKU}.
Following the manipulation and simplification steps as in \eqref{eq_CDF_GAMMA_S_OS_BHKU} and \eqref{eq_CDF_Gamma_S_OS_BHKA_psi_form}, we obtain
\begin{align}\label{eq_Gamma_S_BHKU_OS_High_SNR}
 F_{\hat{\Gamma}_{S}^{(k^*)}}(x) 
  &=1-\sum _{k=1}^K  \sum_{\lb(\mathbf{m_D},\mathbf{n},\mathbf{m_E}\rb)\in\mathcal{X}} (-1)^{k+1} \binom{K}{k} \zeta \nn\\
  &\times \Upsilon \frac{x^{\widehat{m}_{D}^{(k)}}}{\left(x+\frac{(n^{(q)}+1)\lambda_D}{\lambda_E}\right)^{\beta^{(q)}}},
\end{align}
with the definition of $\Upsilon$ as in \eqref{eq_upsilon_OSH}. In \eqref{eq_Gamma_S_BHKU_OS_High_SNR}, we denote by $\mathcal{X}$ the set of vector tuples $\lb(\mathbf{m_D},\mathbf{n},\mathbf{m_E}\rb)$ and the definitions of $\mathbf{m_D},\mathbf{n}$, and $\mathbf{m_E}$ same as in \eqref{eq_CDF_Gamma_S_OS_BHKA_psi_form}.  
Substituting  \eqref{eq_Gamma_S_BHKU_OS_High_SNR} in \eqref{eq_ESR_CDF_eqn}, the high-SNR ESR is evaluated as
\begin{align}\label{eq_C_erg_OS_BHKU_High_SNR_int_form}
 C_{erg} &=\frac{1}{\ln(2)} \sum_{k=1}^K \sum_{\lb(\mathbf{m_D},\mathbf{n},\mathbf{m_E}\rb)\in\mathcal{X}} (-1)^{k+1} \binom{K}{k} \zeta \Upsilon\nn\\
 &\times \big[\mathcal{I}_{0}^{(k)}+\mathcal{I}_{1}^{(k)}\big],
\end{align}
where the solution of $\mathcal{I}_{0}^{(k)}$ is the same as in \eqref{eq_I_0_High_SNR_OS_BHKA_sol}, and the solution of $\mathcal{I}_{1}^{(k)}$ is the same as in \eqref{eq_I_1_High_SNR_OS_BHKA_case_1_sol}-\eqref{eq_I_1_High_SNR_OS_BHKA_case_2_sol}. 
 Further, the approximation $1+\frac{(n^{(q)}+1)\lambda_{D}}{\lambda_{E}} \approx \frac{(n^{(q)}+1)\lambda_{D}}{\lambda_{E}}$ in \eqref{eq_I_0_High_SNR_OS_BHKA_sol}-\eqref{eq_I_1_High_SNR_OS_BHKA_case_2_sol} leads to simplified $\mathcal{I}_{0}^{(k)}$ in \eqref{eq_I_0_Asympt_OS_BHKA} and  $\mathcal{I}_{1}^{(k)}$ in \eqref{eq_I_1_Asympt_OS_BHKA_case_1}-\eqref{eq_I_1_Asympt_OS_BHKA_case_2}, and finally, $C_{erg}^{\infty}$. 

In Sections \ref{section_sop_ka_scenario}-\ref{section_esr_ku}, we showed that by utilizing the ratio of SNRs in \eqref{eq_Gamma_S_k_n_gen} and its CDF in \eqref{eq_sop_to_CDF_conversion}, the secrecy performance metrics the SOP and ESR can be evaluated in a unified manner.

\section{Results}
This section plots the analytical results and the corresponding simulations. In all the figures, $R_{\text{th}}$ is kept equal to 1.
\begin{figure}
\centering
\begin{subfigure}[b]{0.5\textwidth}
 \centering
    \includegraphics[width=\textwidth]{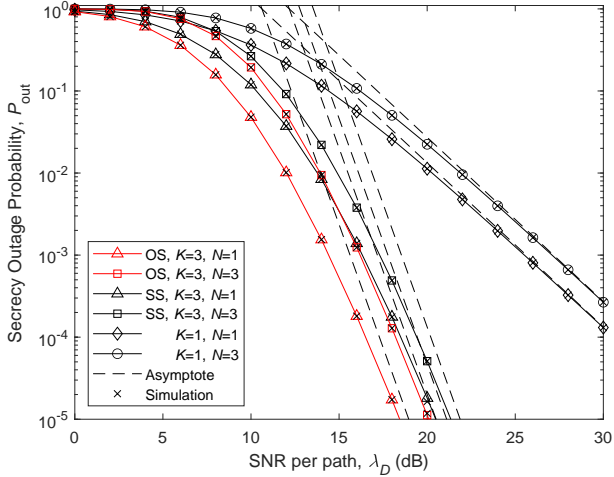} 
 \caption{Variation in $K$ and $N$. $M_D=M_E=2$.}
 \label{fig_SOP_vs_SNR_OS_SS_zeta_1_variation_in_K_N_SNRE_5dB}
\end{subfigure}
\begin{subfigure}[b]{0.5\textwidth}
 \centering
    \includegraphics[width=\textwidth]{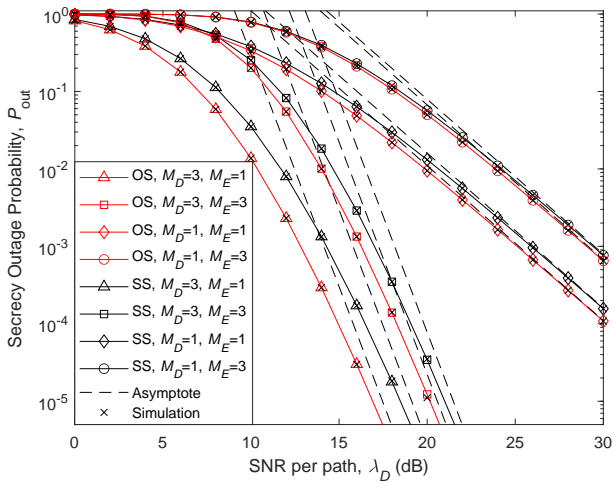}
 \caption{Variation in $M_D$ and $M_E$. $K=N=2$.}
 \label{fig_SOP_vs_SNR_OS_SS_zeta_1_variation_in_MD_ME_SNRE_5dB}
\end{subfigure}
\caption{SOP vs. $\lambda_D$. $\lambda_E=5$ dB.}
\label{fig_SOP_vs_SNR_OS_SS_zeta_1_variation_in_K_N_MD_ME_SNRE_5dB}
\end{figure}
Figs. \ref{fig_SOP_vs_SNR_OS_SS_zeta_1_variation_in_K_N_MD_ME_SNRE_5dB}-\ref{fig_ESR_vs_SNR_OS_SS_zeta_1_variation_in_MD_ME_SNR_SE_9dB} show the secrecy performance when all backhauls are active, i.e., $\zeta=1$. Whereas, 
Figs. \ref{fig_SOP_vs_SNR_OS_SS_BHKA_BHKU_zeta_0_9_variation_in_K_N}-\ref{fig_ESR_HSNR_ASYMP_vs_SNR_OS_variation_in_zeta_K_N} show the secrecy performance under unreliable backhauls, i.e., $\zeta<1$. The high-SNR ESR curves are represented by `-.' and the asymptotes of the SOP and the ESR are represented by `- -'. 
\subsection{All backhaul links active ($\zeta=1$)}
In Fig. \ref{fig_SOP_vs_SNR_OS_SS_zeta_1_variation_in_K_N_SNRE_5dB}, we show the effect of varying $K$ and $N$ on the SOP. The exact SOP curves in the high-SNR regime approach the curves corresponding to the asymptotic SOP. We observe that while increasing $K$ improves the SOP,  increasing $N$ degrades it in both the SS and the OS schemes. However, when 
$\{K,N\}$ are raised equally from $\{1,1\}$ to $\{3,3\}$; there is a net improvement in the SOP performance. It indicates that the number of transmitters impacts the SOP performance more than the number of eavesdroppers. In Fig. \ref{fig_SOP_vs_SNR_OS_SS_zeta_1_variation_in_MD_ME_SNRE_5dB}, the SOP performance is evaluated by varying $M_D$ and $M_E$. As the increase in $M_D$ improves the SNR at the destination, SOP also improves. An increase in $M_E$ improves SNR at the eavesdroppers and thus degrades SOP. The net improvement in the SOP when  $\{M_D,M_E\}$ is increased from $\{1,1\}$ to $\{3,3\}$ shows that $M_D$ has a greater influence on the SOP performance than $M_E$ in both the selection schemes. From the combined observation of Fig. \ref{fig_SOP_vs_SNR_OS_SS_zeta_1_variation_in_K_N_SNRE_5dB} and Fig \ref{fig_SOP_vs_SNR_OS_SS_zeta_1_variation_in_MD_ME_SNRE_5dB}, we note that the number of transmitters and multipath components in the destination channel have more influence on the SOP performance than the number of eavesdroppers and multipath components in the eavesdropper channel. This correlates to our findings in Section \ref{section_div_order_ss} and \ref{section_div_order_os}, where we found that the secrecy diversity order is only dependent on the product  $KM_D$. 

\begin{figure} 
\centering 
\includegraphics[width=3.7in]{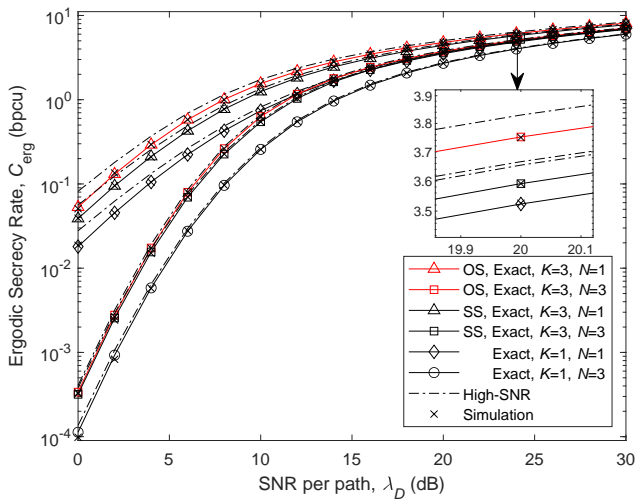} 
\caption{ESR vs. $\lambda_D$ with varying $K$ and $N$. $\zeta=1$, $M_D=M_E=2$ and $\lambda_E=9$ dB.} 
\label{fig_ESR_vs_SNR_OS_SS_zeta_1_variation_in_K_N_SNR_SE_9dB}
\end{figure}

\begin{figure} 
\centering 
\includegraphics[width=3.7in]{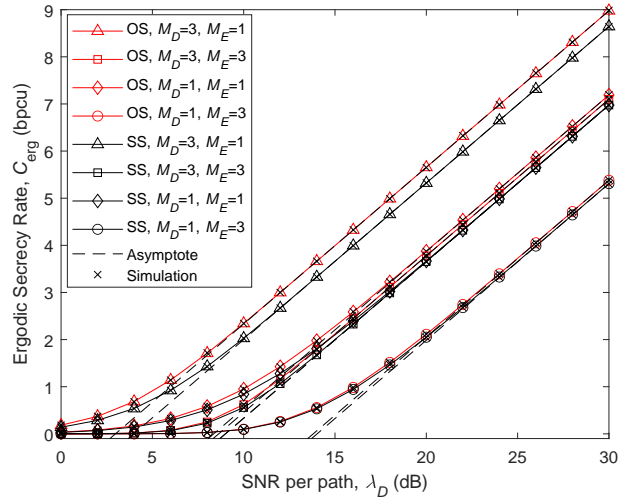} 
\caption{High-SNR ESR vs. $\lambda_D$ with varying $M_D$ and $M_E$. $\zeta=1$, $K=N=2$ and $\lambda_E=9$ dB.} 
\label{fig_ESR_vs_SNR_OS_SS_zeta_1_variation_in_MD_ME_SNR_SE_9dB}
\end{figure}

In Fig. \ref{fig_ESR_vs_SNR_OS_SS_zeta_1_variation_in_K_N_SNR_SE_9dB}, we show the effect of varying $K$ and $N$ on the ESR performance. The high-SNR ESR curves are very close to the corresponding exact ESR curves. 
It shows that the high-SNR ESR expressions derived in Sections \ref{subsection_hsnr_esr_ss_ka} and \ref{subsection_hsnr_esr_os_ka}, although simpler than the exact expressions derived in Sections \ref{subsection_esr_ss_ka} and \ref{subsection_esr_os_ka}, provide a fair approximation of the exact ESR performance.  
Further, we notice that increasing $\{K, N\}$ from $\{1,1\}$ to $\{3,3\}$ degrades the ESR in the low-SNR regime, whereas it improves ESR in the high-SNR regime. It suggests that the effect of transmitter diversity as a result of selection is more pronounced in the high-SNR regime.

In Fig. \ref{fig_ESR_vs_SNR_OS_SS_zeta_1_variation_in_MD_ME_SNR_SE_9dB}, the high-SNR ESR along with corresponding asymptotes are plotted in a linear scale along  the y-axis. We show the effect of varying $M_D$ and $M_E$ on the ESR performance. Increasing $M_D$ by keeping $M_E$ fixed improves ESR, whereas, increasing $M_E$ while keeping $M_D$ fixed degrades the same in both selection schemes. 
Furthermore, increasing $\{M_D,M_E\}$ from $\{1,1\}$ to $\{3,3\}$ degrades the asymptotic ESR in the OS scheme, however, the performance remains same in the SS scheme. This observation is in contrast to that for the SOP in Fig. \ref{fig_SOP_vs_SNR_OS_SS_zeta_1_variation_in_MD_ME_SNRE_5dB}, where there is net improvement in the SOP with the equal increase in both $M_D$ and $M_E$ in both the selection schemes. It shows that the ESR performance is effected by $M_D$ and $M_E$ differently in both the selection schemes. 
Also, the high-SNR ESR slope is independent of $M_D$ and $M_E$, whereas the offset of asymptotic ESR varies with both $M_D$ and $M_E$.


\subsection{Unreliable backhaul links ($\zeta<1$)}

\begin{figure}
\centering
\begin{subfigure}[b]{0.5\textwidth}
 \centering
    \includegraphics[width=\textwidth]{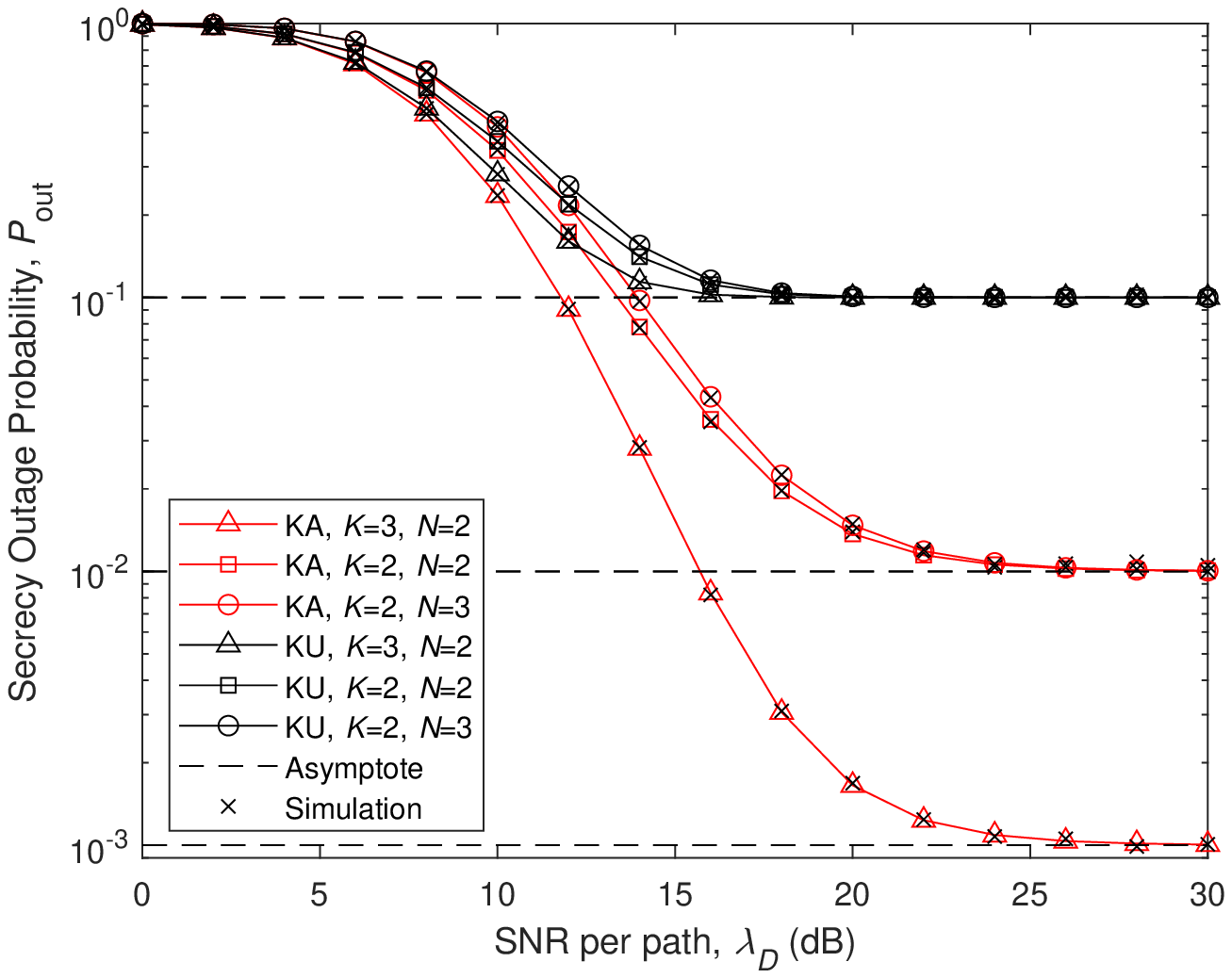}
 \caption{SS scheme} \label{fig_SOP_vs_SNR_SS_BHKA_BHKU_zeta_0_9_variation_in_K_N}
\end{subfigure}
\begin{subfigure}[b]{0.5\textwidth}
 \centering
    \includegraphics[width=\textwidth]{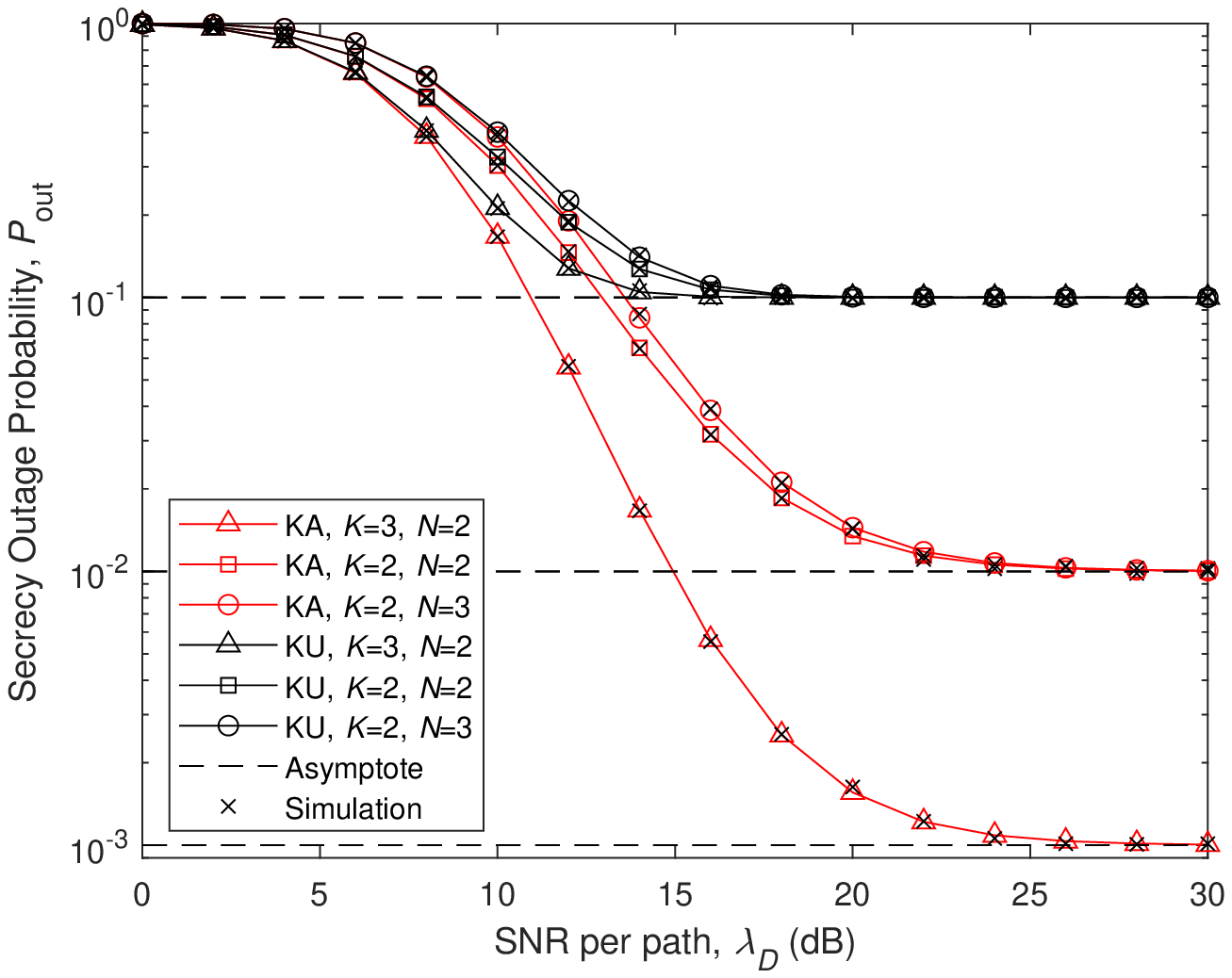} 
 \caption{OS scheme} \label{fig_SOP_vs_SNR_OS_BHKA_BHKU_zeta_0_9_variation_in_K_N}
\end{subfigure}
\caption{SOP vs. $\lambda_D$ with varying $K$ and $N$. $\zeta=0.9$ and $\lambda_E=5$ dB.}
\label{fig_SOP_vs_SNR_OS_SS_BHKA_BHKU_zeta_0_9_variation_in_K_N}
\end{figure}
In Fig. \ref{fig_SOP_vs_SNR_SS_BHKA_BHKU_zeta_0_9_variation_in_K_N} and Fig. \ref{fig_SOP_vs_SNR_OS_BHKA_BHKU_zeta_0_9_variation_in_K_N}, we plot the exact SOP and the corresponding asymptotes by varying $K$ and $N$ under unreliable backhauls.
We observe that in the low SNR regime, the SOP performances in both backhaul KA and KU scenarios are close to each other.  Therefore, the knowledge of backhaul activity does not improve ESR much in both SS and OS schemes.
In the high-SNR regime, the performance in the backhaul KA scenario is better than in the KU scenario. This is because, in the high-SNR regime, the SOP in the backhaul KA scenario is independent of $N$ and depends on $K$ for a given $\zeta$. However, in the backhaul KU scenario, the SOP depends only on $\zeta$.
 This observation is in accordance with the asymptotic analysis performed for the backhaul KA and KU scenarios in Section \ref{subsection_asymp_sop_ss_ka} and \ref{subsection_asymp_sop_ss_ku}, respectively.

\begin{figure}
\centering
\begin{subfigure}[b]{0.5\textwidth}
 \centering
    \includegraphics[width=\textwidth]{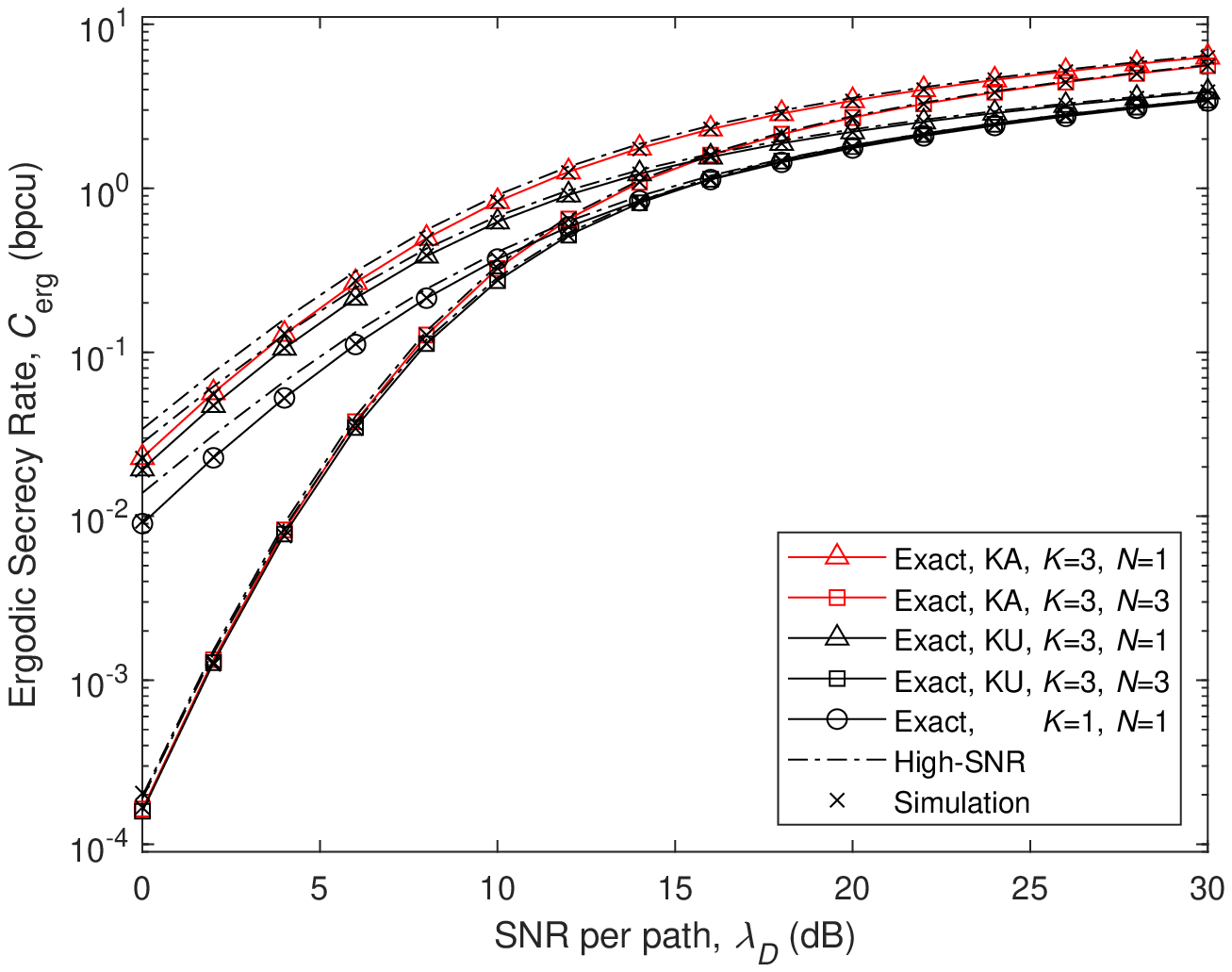}
 \caption{SS scheme}
 \label{fig_ESR_vs_SNR_SS_zeta_5_variation_in_BH_K_N}
\end{subfigure}
\begin{subfigure}[b]{0.5\textwidth}
 \centering
    \includegraphics[width=\textwidth]{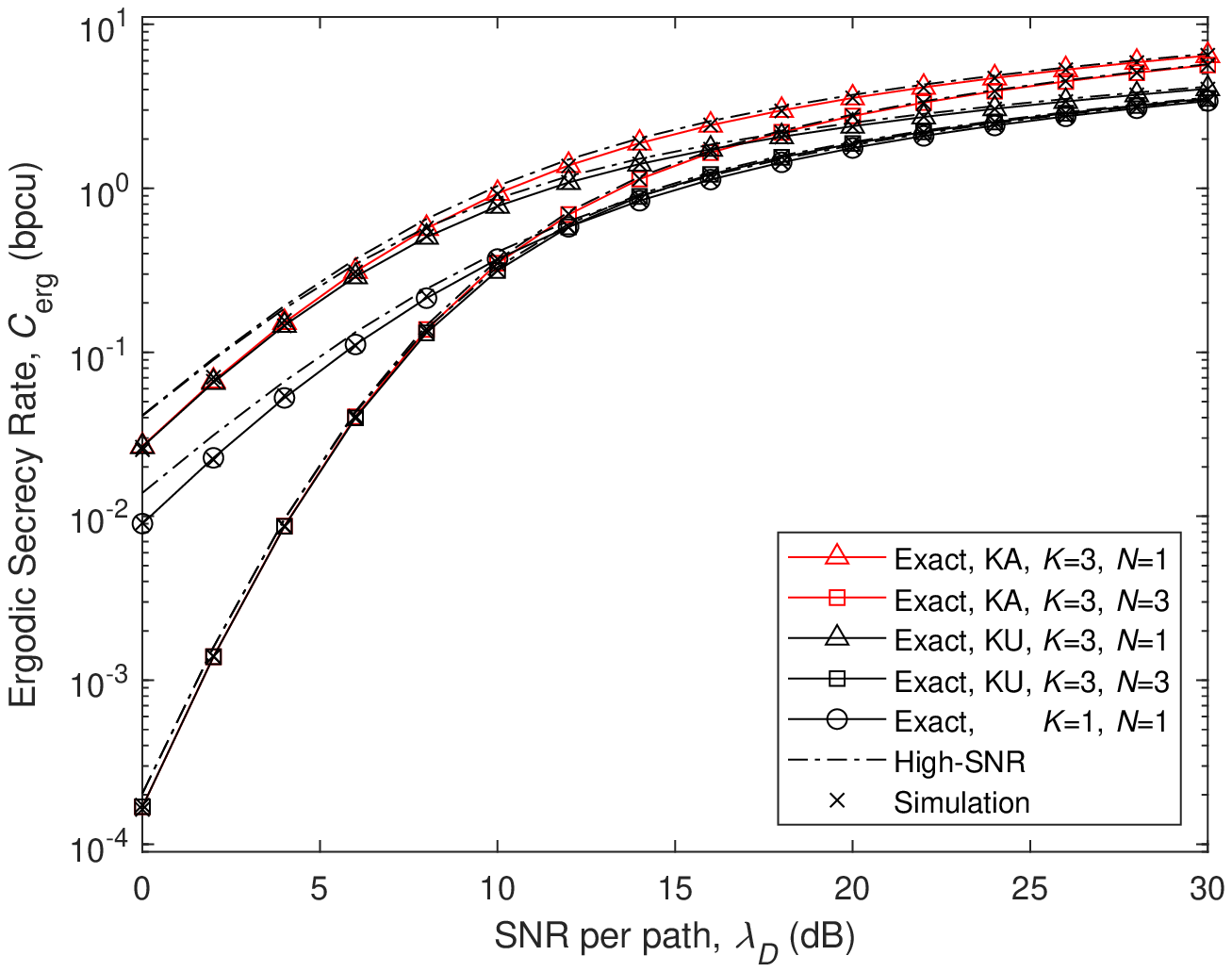} 
 \caption{OS scheme} \label{fig_ESR_vs_SNR_OS_zeta_5_variation_in_BH_K_N}
\end{subfigure}
\caption{ESR vs. $\lambda_D$ with varying $K$ and $N$. $\zeta=0.5$, $M_D=M_E=2$ and $\lambda_E=9$ dB.}
\label{fig_ESR_vs_SNR_OS_SS_variation_in_BH_K_N}
\end{figure}
\begin{figure} 
\centering 
\includegraphics[width=3.7in]{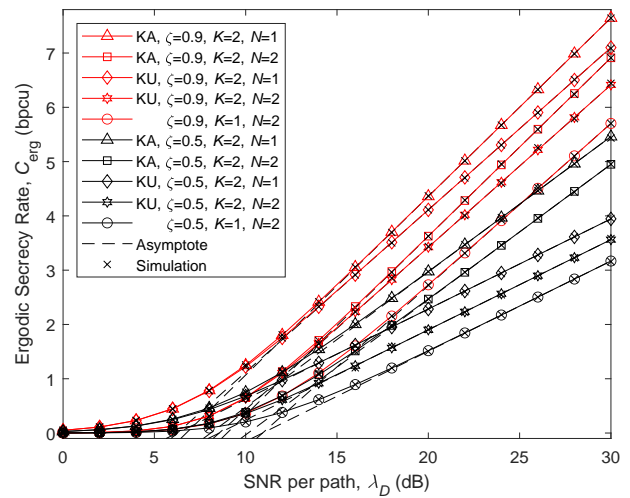} 
\caption{ESR vs. $\lambda_D$ in OS with varying $K$ and $\zeta$. $M_D=M_E=2$ and $\lambda_E=9$ dB.} 
\label{fig_ESR_HSNR_ASYMP_vs_SNR_OS_variation_in_zeta_K_N}
\end{figure}
In Fig. \ref{fig_ESR_vs_SNR_SS_zeta_5_variation_in_BH_K_N} and Fig. \ref{fig_ESR_vs_SNR_OS_zeta_5_variation_in_BH_K_N}, we plot the exact ESR and the corresponding high-SNR ESR under unreliable backhauls by varying $K$ and $N$. 
As in the case of the SOP performance in Fig. \ref{fig_SOP_vs_SNR_SS_BHKA_BHKU_zeta_0_9_variation_in_K_N} and Fig. \ref{fig_SOP_vs_SNR_OS_BHKA_BHKU_zeta_0_9_variation_in_K_N}, the knowledge of backhaul activity does not improve the ESR performance.  
Similar to Fig. \ref{fig_ESR_vs_SNR_OS_SS_zeta_1_variation_in_K_N_SNR_SE_9dB}, varying $\{K, N\}$ from $\{1, 1\}$ to $\{3, 3\}$ degrades the ESR in the low-SNR regime, and improves it in the high-SNR regime. In addition to the observations in Fig. \ref{fig_ESR_vs_SNR_OS_SS_zeta_1_variation_in_K_N_SNR_SE_9dB}, here, we observe that increasing $K$ improves the ESR in the backhaul KA scenario more than in the KU scenario. 

Fig. \ref{fig_ESR_HSNR_ASYMP_vs_SNR_OS_variation_in_zeta_K_N} shows the effect of varying $\zeta$ along with $K$ and $N$ on the high-SNR ESR and the corresponding asymptotes in a linear scale along the vertical axis. Firstly, we observe that the asymptotic curves corresponding to $\zeta=0.9$  cross over the curves for $\zeta=0.5$ in both the backhaul KA and KU scenarios. Also, for the asymptotes with the same slope, the ESR offset varies with the number of eavesdroppers. It indicates that the slope of ESR depends on $\zeta$ in both scenarios. Secondly, for a given $\zeta$ in the backhaul KA scenario, this slope  varies with $K$  and is independent of $N$. In the backhaul KU scenario, it is independent of both $K$ and $N$. These two observations lead to the conclusion that the slope in the backhaul KA scenario depends on the number of transmitters and the backhaul reliability factor, whereas in the KU scenario, it depends on the backhaul reliability factor only. The slope does not vary with the number of eavesdroppers.  Therefore, increasing the number of transmitters is more effective in the backhaul KA than in the KU scenario. 

\section{Conclusion}
In this paper, considering the knowledge available and unavailable scenarios of the backhaul activity, the secrecy performance of SS and OS transmitter selection schemes is investigated. The system is consisting of multiple transmitters, multiple eavesdroppers, and  a single destination, with SC-CP signaling in frequency-selective fading channels. 
We demonstrate that, for secrecy performance, the effect of the number of transmitters is more significant than the number of eavesdroppers. When all backhauls are active, the OS scheme takes better advantage of the favorable conditions, including increased transmitters and the SNR at the destination, than the SS scheme. We show that when backhauls are unreliable, their reliability factor governs the asymptotic SOP limit and the slope of the asymptotic ESR. However, the number of eavesdroppers does not influence these parameters. Increasing the number of transmitters improves both these parameters in the backhaul KA scenario only. The generalized approach to obtain the closed-form SOP and the ESR from the CDF of the ratio of destination to the eavesdropper SNR can also be utilized to evaluate other secrecy metrics.

\bibliographystyle{IEEEtran}
\bibliography{IEEEabrv, ref}
\end{document}